\documentclass{ar-1col}
\usepackage{rotating}%
\usepackage{cite}
\usepackage{amsmath}
\usepackage{amssymb}
\usepackage{soul}
\usepackage{url}
\jname{Xxxx. Xxx. Xxx. Xxx.}
\jvol{00}
\jyear{YYYY}
\doi{10.1146/((please add article doi))}

\begin{document}

\markboth{Riccardo Comin and Andrea Damascelli}{Resonant x-ray scattering studies of charge order in cuprates}

\title{Resonant x-ray scattering studies of charge order in cuprates}

\author{Riccardo Comin${}^{1,2,3}$ and Andrea Damascelli${}^{2,3}$
\affil{$^1$Department of Electrical and Computer Engineering, University of Toronto, Toronto, Ontario M5S 3G4, Canada}
\affil{$^2$Department of Physics {\rm {\&}} Astronomy, University of British Columbia, Vancouver, British Columbia V6T\,1Z1, Canada}
\affil{$^3$Quantum Matter Institute, University of British Columbia, Vancouver, British Columbia V6T\,1Z4, Canada}
\affil{r.comin@utoronto.ca; damascelli@physics.ubc.ca}}


\begin{abstract}
X-ray techniques have been used for more than a century to study the atomic and electronic structure in virtually any type of material. The advent of correlated electron systems, in particular complex oxides, brought about new scientific challenges and opportunities for the advancement of conventional x-ray methods. In this context, the need for new approaches capable of selectively sensing new forms of orders involving all degrees of freedom -- charge, orbital, spin, and lattice -- paved the way for the emergence and success of resonant x-ray scattering, which has become an increasingly popular and powerful tool for the study of electronic ordering phenomena in solids. Here we review the recent resonant x-ray scattering breakthroughs in the copper oxide high-temperature superconductors, in particular regarding the phenomenon of charge order -- a broken-symmetry state occurring when valence electrons self-organize into periodic structures. After a brief historical perspective on charge order, we outline the milestones in the development of resonant x-ray scattering, as well as the basic theoretical formalism underlying its unique capabilities. The rest of the review will focus on the recent contributions of resonant scattering to the tremendous advancements in our description and understanding of charge order. To conclude, we propose a series of present and upcoming challenges, and discuss the future outlook for this technique.
\end{abstract}


\maketitle

\tableofcontents

\section{INTRODUCTION AND BRIEF HISTORICAL SYNOPSIS}

Over the years, transition metal oxides have represented a traditional platform for strongly correlated electron physics, which has nowadays become a field of its own, encompassing several classes of compounds with one common denominator: the localized character of the low-energy electronic wavefunctions (with $d$ or $f$ orbital character) and the corresponding prominence of Coulomb interactions in driving the electronic properties of these materials. In conventional metals or semiconductors, the fermiology is essentially determined by the lowering of the total kinetic energy which becomes possible when a periodic potential supports the delocalization of the local orbitals into extended wavefunctions with a well-defined momentum and a correspondigly homogeneous distribution of the charge density. In correlated electron systems, the on-site Coulomb repulsion between two electrons in the same $d$ or $f$ orbital can overcome the kinetic energy part of the Hamiltonian, inducing the electronic system to find new ways to lower its total energy, often by spontaneous breaking of the native symmetries of the lattice (translational and/or point group symmetry). This tendency leads to the emergence of a rich variety of symmetry-broken electronic phases, and represents a distinctive trademark of strongly correlated systems, spanning across families of compounds otherwise very different from a chemical standpoint \cite{Dagotto2005,Norman2011}.

Within the extended class of correlated electron systems, copper oxides represent a unique breed due to the mixed character of the electronic bands, frustrated by the conflicting interplay between the O-$2p$ states -- promoting itinerancy and electron hopping -- and the Cu-$3d$ states -- which hinder charge fluctuations and are conducive to Mott-Hubbard physics and an insulating, antiferromagnetic (AF) ground state \cite{ZSA}. This delicate balance can be tuned and controlled by carrier doping, resulting in a phase diagram of astonishing richness and complexity, yet to be fully understood \cite{Lee_2006}. Antiferromagnetism has been long known as the ground state in charge-transfer insulating undoped copper oxides \cite{ZSA}, while unconventional superconductivity was discovered in 1986 \cite{Bednorz_1986}. To date, several broken symmetries have been detected in the cuprates, which can be categorized into zero-momentum ($Q \!=\! 0$) and finite-momentum ($Q \!\neq\! 0$) orders, breaking rotational and translational symmetry, respectively. These different types of order can involve both the charge sector (nematic state for $Q \!=\! 0$, and charge-density-wave for $Q \!\neq\! 0$) and the spin sector ($Q \!=\! 0$ magnetic order, and spin-density-wave or antiferromagnetism with $Q \!\neq\! 0$), leading to several proposed forms of intertwined electronic orders \cite{Emery1999,Kivelson_RMP,Sachdev2004,birgeneau2006,Varma2006,Li2008,Berg2009,Norman2011,Davis_Lee_2013,Fradkin_2015}. 
\begin{marginnote}
\entry{AF}{Antiferromagnetic}
\entry{HTSC}{High-temperature superconductor}
\end{marginnote}

\subsection{First observations of charge order in cuprates}

\begin{figure}[t!]
\includegraphics[width=1\linewidth]{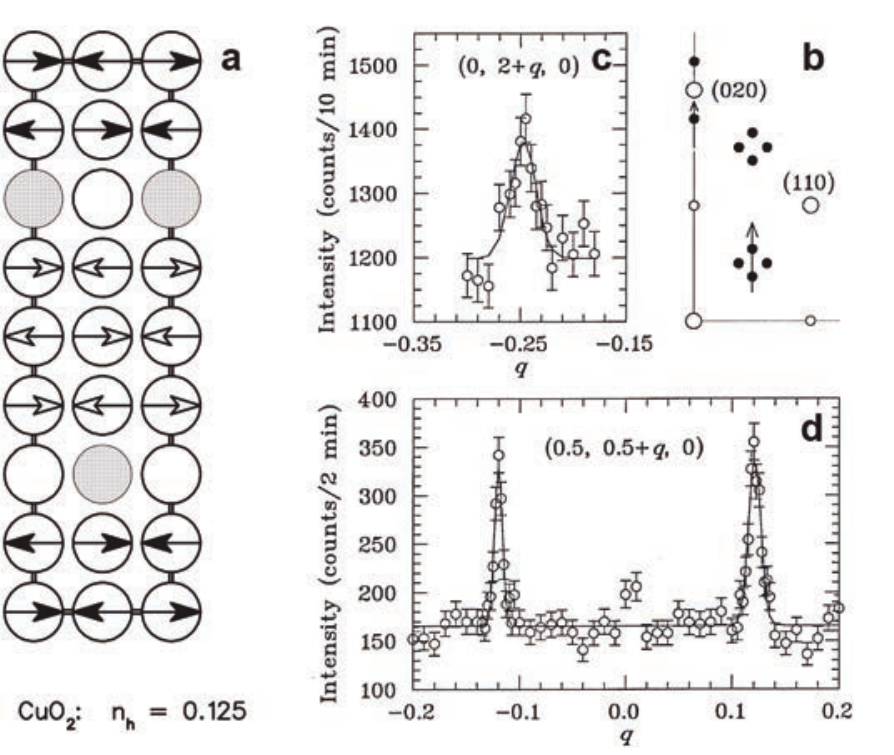}
\caption{\textbf{Neutron scattering discovery of stripe order in 12\,\% doped La${}_{1.6-x}$Nd${}_{0.4}$Sr${}_{x}$Cu${}_{2}$O${}_{4}$.} (\textit{a}) Schematic representation of the stripe pattern: circles represent the Cu sites in the CuO${}_{2}$ plane, with arrows denoting the Cu spins and gray circles indicating the location of the doped holes. (\textit{b}) $(HK0)$ projection of the reciprocal space showing the location of the measured charge and spin satellites (black circles) along the $(010)$ and $(110)$ directions, respectively; white circles represent Bragg reflections. (\textit{c,d}) Elastic neutron scattering measurement of the charge order peak at $(0,1.75,0)$ (\textit{c}) and of the magnetic peaks at $(0.5,0.375,0)$ and $(0.5,0.625,0)$ (\textit{d}), with scans along the arrows marked in (\textit{b}). Readapted from Ref.\,\citen{tranquada1995}.}.
\label{Intro_fig}
\end{figure}

In this review we will focus on charge order, which is defined as an electronic phase breaking translational symmetry via a self-organization of the electrons into periodic structures incompatible with the periodicity of the underlying lattice. The denomination of \textit{charge order}, or equivalently \textit{charge-density-wave}, has over time embraced a broad phenomenology. However, charge order was first discovered in the form of stripe order, whose real-space representation is depicted in Fig.\,\ref{Intro_fig}a. In this ordered state, the doped holes (dark circles) are segregated into unidirectional structures which act as boundaries separating undoped domains characterized by AF order. Therefore, stripe order in its generalized meaning is an electronic ground state characterized by a \textit{combination of magnetic order and charge order with specific geometrical constraints on the ordering wavevectors}.

Historically, the first proof of stripe order came from neutron scattering, a momentum-resolved technique which had been extensively used in the early days of high-temperature superconductors (HTSCs) due to its excellent energy resolution and large magnetic cross section. The first neutron scattering studies of HTSCs focused on the doping and temperature dependence of AF order in YBa${}_{2}$Cu${}_{3}$O${}_{6+x}$ (YBCO) \cite{tranquada1988} and La${}_{2-x}$Sr${}_{x}$CuO${}_{4}$ (LSCO) \cite{Birgeneau_1988}, which was determined to be located at wavevectors ${\mathbf{Q}}_{\mathrm{AF}} \!=\! \left( \pm 1/2, \pm 1/2, L \right)$\footnote{Hereafter we will use reciprocal lattice units (r.l.u.) for wavevectors in momentum space. Reciprocal lattice units represent a notation where wavevectors are expressed as $\mathbf{Q} \!=\! \left( H, K, L \right)$, corresponding to $\mathbf{Q} \!=\! \left( H \cdot 2 \pi /a , K \cdot 2 \pi /b , L \cdot 2 \pi /c \right)$ in physical units (typically, ${\AA}^{-1}$). Also, unless otherwise stated, we will refer the wavevectors to the undistorted unit cell, where the \textbf{a} and \textbf{b} axes (and correspondingly the reciprocal axes \textbf{H} and \textbf{K}) are parallel to the Cu-O bond directions with lattice parameters equal to the nearest Cu-Cu distance.}. Around the same time, the first experimental hint of stripe order was revealed by the independent neutron scattering measurements of Yoshizawa \textit{et al.} \cite{Yoshizawa_1988} and Birgeneau \textit{et al.} \cite{Birgeneau_1989}, who reported incommensurate AF modulations in underdoped La${}_{1.92}$Sr${}_{0.08}$CuO${}_{4}$ (8\,\% hole doping) and La${}_{1.89}$Sr${}_{0.11}$CuO${}_{4}$ (11\,\% hole doping), respectively, suggesting the presence of an electronic phase characterized by the quasi-static ordering of the doped holes. A subsequent study by Thurston \textit{et al.} \cite{Thurston_1989} revealed that the incommensurate AF order could only be found in the superconducting phase of LSCO [it was later shown to be present also in the insulating phase, albeit with different modulation vector \cite{Fujita_2002_PRB}]. The full momentum structure and doping dependence of incommensurate spin order in LSCO was later uncovered by Cheong \textit{et al.} \cite{Cheong_1991}, and shown to manifest itself as a set of four magnetic peaks at ${\mathbf{Q}}_{\mathrm{AF-IC}} \!=\! \left( 1/2 \pm {\delta}_{\mathrm{IC}}, 1/2, L \right)$ and ${\mathbf{Q}}_{\mathrm{AF-IC}} \!=\! \left( 1/2 , 1/2 \pm {\delta}_{\mathrm{IC}}, L \right)$ (see diagram in Fig.\,\ref{Intro_fig}b), with ${\delta}_{\mathrm{IC}}$ representing the doping-dependent incommensurability.
\begin{marginnote}
\entry{YBCO}{YBa${}_{2}$Cu${}_{3}$O${}_{6+x}$}
\entry{LSCO}{La${}_{2-x}$Sr${}_{x}$CuO${}_{4}$}
\entry{LNSCO}{La${}_{2-x-y}$Nd${}_{y}$Sr${}_{x}$CuO${}_{4+\delta}$}
\end{marginnote}

The experimental observations from neutron scattering spurred an intense theoretical activity, aimed at addressing two primary phenomenological findings: (i) the momentum structure of the incommensurate AF order and its relationship to the ordering of the doped holes; and (ii) the emergence of incommensurate AF order only within the superconducting phase, which pointed at an intimate interplay between short-range magnetism and superconductivity in the cuprates. The short-range attractive interaction between segregated charges was initially proposed as a pairing mechanism for the superconducting state \cite{Schrieffer_1988}, while several numerical studies of the Hubbard model near half-filling were performed under different conditions: using mean-field theory without \cite{Machida1989,Castellani1995} and with \cite{Zaanen1989} charge fluctuations; within Hartree-Fock approximation \cite{Poilblanc1989,Schulz_1990}; and by using exact diagonalization methods \cite{Emery1990}. All these studies pointed to a symmetry-broken ground state with short-ranged charge inhomogeneities organized into an ordered pattern of unidirectional charged arrays, segregated away from the magnetic domains and acting as boundaries for the latter \cite{Zaanen1989}. The stripe structure with holes condensed along rivers of charge-separating antiferromagnetic domains was expected to manifest itself also as a periodic modulation of the charge, with a wavevector twice as long as in the case of the incommensurate AF peaks \cite{Machida1989,Zaanen1989} and tied to the hole concentration (``a counting rule stating that the number of domain line unit cells is equal to the number of carriers.", from Ref.\,\citen{Zaanen1989}).

Such a stripe state was found shortly thereafter by Tranquada \textit{et al.} in the nickelate compound La${}_{2}$NiO${}_{4+\delta}$ \cite{Tranquada_1994_LSNO}, as evidenced by the presence of charge\footnote{Neutrons can probe charge order indirectly, by detecting the associated distortion of the underlying lattice.} and magnetic satellite reflections in neutron scattering at the (collinear) wavevectors ${\mathbf{Q}}_{\mathrm{charge}} \!=\! \left( H \pm 2 \epsilon, K \pm 2 \epsilon, L \right)$ and ${\mathbf{Q}}_{\mathrm{spin}} \!=\! \left( H \pm \epsilon, K \pm \epsilon, L \right)$, respectively, reflecting the fundamental relationship between the spin and charge ordering incommensurability factors: ${\delta}_{\mathrm{charge}} \!=\! 2 {\delta}_{\mathrm{spin}}$. These two characteristics (collinearity and two-fold proportionality of wavevectors) represent the basic conditions for stripe order, as opposed to other geometries (such as checkerboard). The search for static stripe order in the isostructural LSCO cuprate compound was complicated by the necessity for native fourfold symmetry in the CuO${}_{2}$ planes to be broken in order to accommodate a stripe state. A solution was eventually obtained with the addition of rare earth impurities (Nd and Eu, in place of La) distorting the lattice in the otherwise square-symmetric CuO${}_{2}$ planes and providing an inner strain field stabilizing stripe order. Consequently, in 1995 Tranquada \textit{et al.} \cite{tranquada1995} used neutron scattering to detect charge and spin order peaks in the same sample of underdoped Nd-substituted LSCO, La${}_{2-x-y}$Nd${}_{y}$Sr${}_{x}$Cu${}_{2}$O${}_{4+\delta}$ (LNSCO) at ${\mathbf{Q}}_{\mathrm{charge}} \!=\! \left( 2, 2 \pm 0.25, 0 \right)$ and ${\mathbf{Q}}_{\mathrm{spin}} \!=\! \left( 0.5, 0.5 \pm 0.125, 0 \right)$, as shown in Fig.\,\ref{Intro_fig}c and d, respectively. This finding confirmed the wavevector relationship already proven in the nickelate compound, namely that ${\delta}_{\mathrm{charge}} \!=\! 0.25 \!=\! 2 \cdot 0.125 \!=\! 2 {\delta}_{\mathrm{spin}}$, as well as the equivalence between the incommensurability and the doping -- with the former amounting to $\sim\! 1/8 \!=\! 0.125$ at the `magical' doping level of 12.5\,\% -- which is another signature of stripe order. This work represented the first direct evidence of stripe order in the cuprates and, together with closely following studies using again neutron \cite{Tranquada_1996,Fujita_2002}, as well as x-ray \cite{vZimmermann_1998} scattering, initiated the whole experimental field of charge ordering phenomena in HTSCs.

For more extended reviews of neutron scattering studies in the cuprates, we refer the reader to Refs.\,\citen{Kastner_review_1998,birgeneau2006}, while for a comprehensive discussion of the theoretical works on charge and spin order we refer the reader to Refs.\,\citen{Emery1999,Kivelson_RMP,Fradkin_2015}.

\subsection{Imaging charge order in real space}

Towards the end of the 90's, a series of advancements in scanning tunnelling microscopy (STM) methods made it possible to obtain atomically-resolved maps with spectroscopic information on the energy-dependent local density of states (LDOS) via the measurement of the bias-dependent tip-sample differential tunnelling conductance as a function of the spatial coordinate $\mathbf{r}$, $dI / dV = g \left( \mathbf{r}, V \right) \propto LDOS \left( \mathbf{r}, E \!=\! eV \right) $. The atomic resolution enabled by STM setups was key to the detection of density modulations with remarkably short correlation lengths, leading the way to a new era in the study of the nanoscale interplay between different electronic orders and of the resulting granular structure in HTSCs.
\begin{marginnote}
\entry{STM}{Scanning Tunnelling Microscopy}
\entry{Bi2212}{Bi${}_{2}$Sr${}_{2}$CaCu${}_{2}$O${}_{8+\delta}$}
\entry{Bi2201}{Bi${}_{2−y}$Pb${}_{y}$Sr${}_{2-z}$La${}_{z}$CuO${}_{6+\delta}$}
\entry{Na-CCOC}{Na${}_{x}$Ca${}_{2-x}$CuO${}_{2}$Cl${}_{2}$}
\end{marginnote}
\begin{figure}[t!]
\includegraphics[width=1\linewidth]{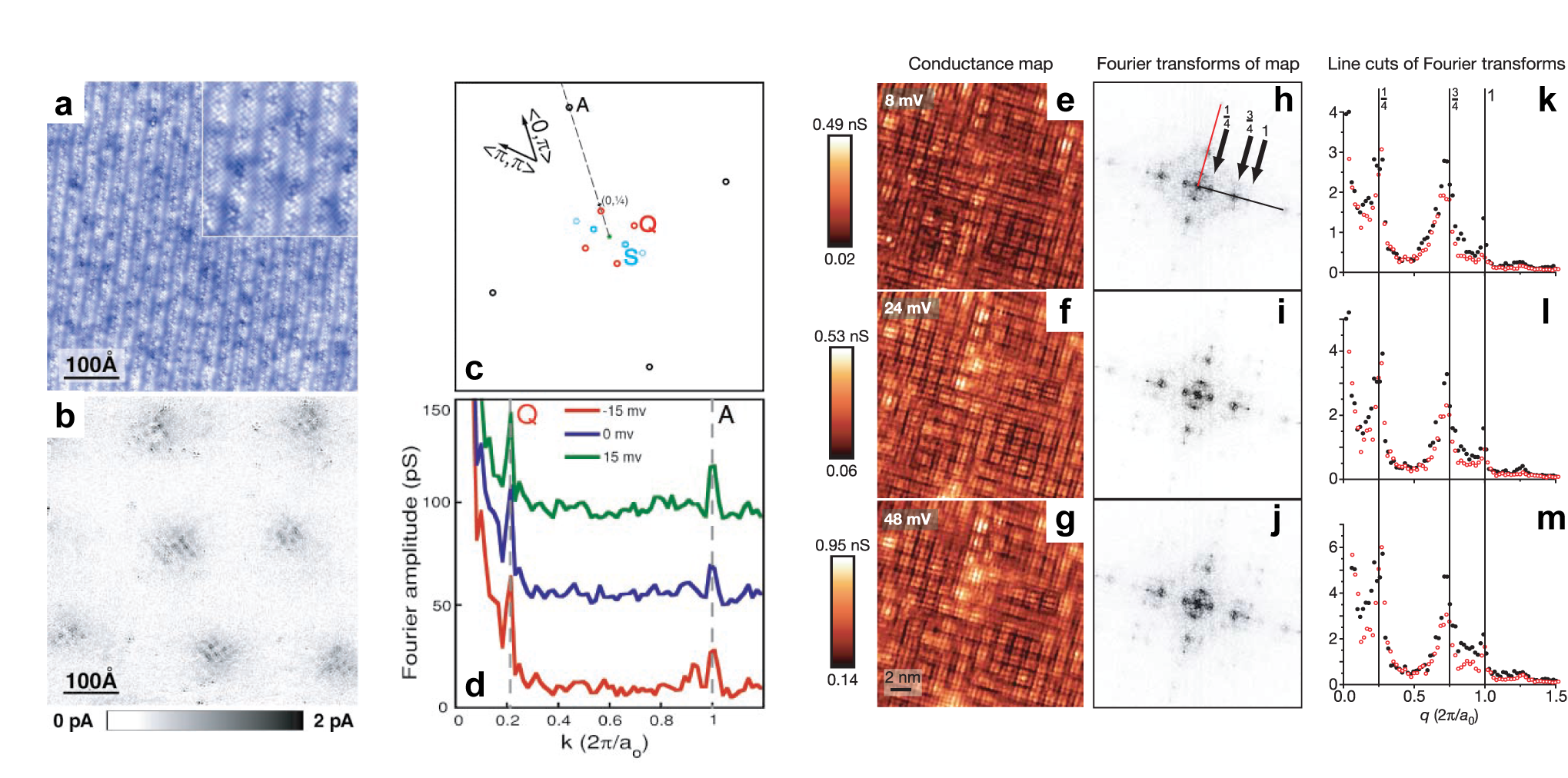}
\caption{\textbf{Scanning tunnelling microscopy explorations of charge order in Bi-based cuprates.} (\textit{a}) Low-temperature topographic map of a cleaved surface of slightly overdoped (${T}_{\mathrm{c}} \!=\! 89$\,K) Bi${}_{2}$Sr${}_{2}$CaCu${}_{2}$O${}_{8+\delta}$ (Bi2212). (\textit{b}) Differential tunnelling conductance ($dI / dV$) map (integrated between 1 and 12\,meV) between an applied magnetic field of 5 and 0\,T, showing the appearance of period-4 density modulations within the magnetic vortex cores, and over the same spatial region of (\textit{a}). Readapted from Ref.\,\citen{hoffman2002}. (\textit{c}) Momentum space representation of the periodic structures observed using STM on Bi2212, including the Fourier peaks from the atomic lattice (A) and the structural supermodulation (S), as well as the charge order peaks (Q) at $\mathbf{Q} \sim (\pm 0.21,0)$ and $\mathbf{Q} \sim (0, \pm 0.21)$, in reciprocal lattice units. (\textit{d}) Energy-resolved Fourier-transformed conductance map along the $(10)$ direction, showing no dependence of the charge order peak position (Q) on energy, thereby demonstrating the independence of this phenomenon from quasiparticle interference effects. Readapted from Ref.\,\citen{vershinin2004}. (\textit{e-g}) Conductance maps in Na-CCOC at different bias voltages, and (\textit{h-j}) corresponding 2D Fourier transforms, whose linecuts (\textit{k-m}) along the Cu-O bond directions show the charge order spatial frequencies and the lack of energy dispersion in the momentum-resolved structures. Readapted from Ref.\,\citen{hanaguri2004}.}\label{STM_fig}
\end{figure}

The new spectroscopic imaging scanning tunnelling microscopy (SI-STM) capabilities were soon applied to the study of HTSCs, with the family of choice being that of Bi-based cuprates, owing to the presence of natural cleavage planes yielding extended, atomically-flat surfaces that turned out to be ideal for STM studies. The first compound to be studied was Bi${}_{2}$Sr${}_{2}$CaCu${}_{2}$O${}_{8+\delta}$ (Bi2212), a layered cuprate with a BiO--SrO--CuO${}_{2}$--Ca--CuO${}_{2}$--SrO--BiO stacking of crystallographic planes, characterized by a weak, van der Waals-type bonding between adjacent BiO layers. In 2002, 7 years after the original discovery of stripe order in the La-based compounds, the STM investigation of Bi2212 by Hoffman \textit{et al.} \cite{hoffman2002} revealed the presence charge order in yet another cuprate family. The authors studied cleaved surfaces of slightly overdoped Bi2212 samples, whose topographic STM map is shown in Fig.\,\ref{STM_fig}a. The corresponding differential conductance map over the same field of view, and upon application of a 5\,T magnetic field, is depicted in Fig.\,\ref{STM_fig}b. In the field-induced vortex cores (darker spots), where the superconducting order is suppressed, four-fold, bi-directional modulations of the local density of states can be clearly imaged. This density modulation, with correlation lengths of the order of 30\,\AA, provided experimental evidence in support of the expectation for a charge instability \cite{Chen_2002} in the doping range where the field-induced period-8 spin density modulation was discovered in LSCO shortly before \cite{Lake2001}. The emergence of the periodic modulations within the vortex cores further suggested a direct competition between charge order and superconductivity, in analogy to the phenomenology in the La-based cuprates \cite{Tranquada_1997}.

While similar evidence for charge order was found by Howald \textit{et al.} \cite{howald2003} at low temperature and in the absence of a magnetic field, around the same time the discovery of quasi-particle interference (QPI) \cite{Hoffman_QPI} in the superconducting state and below an energy scale of the order of the superconducting gap underscored the importance of spectroscopic measurements resolving the observed structure as a function of sample-to-tip bias voltage. The energy dependence of the charge order signal in Fourier space was later explored in the STM study of Vershinin \textit{et al.} on Bi2212 \cite{vershinin2004} and Hanaguri \textit{et al.} \cite{hanaguri2004} on Na-doped Ca${}_{2}$CuO${}_{2}$Cl${}_{2}$ (Na-CCOC). Figure \ref{STM_fig}c shows the reciprocal space chart of Bi2212 -- obtained by Fourier transforming the differential tunnelling conductance map -- in the pseudogap state ($T \!=\! 100$\,K), where QPI effects from superconducting quasiparticles are not active. The two-dimensional momentum structure of charge order can be also followed as a function of energy (corresponding to the bias between the STM tip and the underlying surface), and Fig.\,\ref{STM_fig}d shows a series of linecuts across the charge ordering wavevector ${\mathbf{Q}} \!\sim\! 0.21$\,r.l.u., along the $(H0)$ direction, as a function of energy. A non-dispersive charge order peak (Q) was found to be present across an extended energy range with almost constant amplitude, thereby proving its independence from the energy-dependent features due to quasi-particle interference. Charge order in highly underdoped ($0.08 \!<\! p \!<\! 0.12$) Na-CCOC, including non-superconducting samples, was discovered around the same period and shown to similarly exhibit an energy-independent momentum structure across a broad range of energies in the pseudogap state. Panels e-g and h-j in Fig.\,\ref{STM_fig} show, respectively, the two-dimensional real- and reciprocal- (Fourier-transformed) maps of the differential tunnelling conductance at 8, 24, and 48\,mV bias. Distinctive features can be identified in the Fourier transform maps that correspond to near period-4 electronic modulations along the Cu-O bond directions. The energy-independence of these features appears evident from the linecuts shown in Fig.\,\ref{STM_fig}k-m, demonstrating the presence of a broken symmetry driving static electronic modulations affecting all the electronic states over an extended energy range.

The charge-ordered state in Bi2212 and Na-CCOC was later shown by Kohsaka \textit{et al.} \cite{koshaka2007} to break down into bond-centered, locally unidirectional domains, as revealed by tunnelling asymmetry imaging. These measurements also revealed the prominence of these electronic modulations at large energy scales, of the order of the pseudogap energy, a tendency confirmed in subsequent studies \cite{Koshaka2008,Lee2009}. Evidence for charge order in single-layered Bi-based compounds, Bi${}_{2−y}$Pb${}_{y}$Sr${}_{2-z}$La${}_{z}$CuO${}_{6+\delta}$ (Bi2201), was reported in 2008 by Wise \textit{et al.} \cite{wise2008} over an extended range of dopings, demonstrating that the charge order wavevector evolution is compatible with a possible instability arising from the antinodal region of the Fermi surface. A similar doping dependence was also detected in Bi2212 \cite{mesaros2011,Fujita2011,dSN_Science}. In more recent years, a possible connection between charge order and the pseudogap state has been unveiled in Bi2212 using high-temperature STM to correlate the onset of the electronic signal from incipient stripes with the pseudogap temperature ${T}^{*}$ \cite{parker2010,daSilvaNeto}. These findings were followed by temperature-dependent measurements which show a competition between charge order and superconductivity near the Fermi energy \cite{dSN_Science}. Even more recently, it was shown that the zero-temperature density-wave order disappears in the vicinity of a putative quantum critical point located at a doping of $p \!\sim\! 0.16$ for Bi2201 \cite{He2014,Peli_2015} and $p \!\sim\! 0.19$ for Bi2212 \cite{Fujita2014_QCP}. Throughout the years, STM and SI-STM provided a constant stream of scientific results and valuable insights on the existence, nature, symmetry, and nanoscale structure of charge order in underdoped cuprates; many of these studies have now been covered by several reviews \cite{Renner_RMP,Fujita2015_Springer,dSN2015_ARCMP}.
 
\section{RESONANT X-RAY METHODS}

\subsection{Resonant x-ray scattering in a nutshell}

X-rays have been long used to investigate the inner structure of matter, thanks to the interaction of light with the electronic clouds surrounding the atomic nuclei \cite{Nielsen_2011}. The role of photon energy has traditionally been secondary -- with the exception of anomalous x-ray diffraction, a technique which relies on the strongly varying x-ray absorption near the absorption edges of certain elements to simplify the phase retrieval problem in crystallography and provide a full reconstruction of the atomic positions in complex macromolecular systems \cite{Hendrickson_2014}. The first glimpses of resonant x-ray effects were revealed at the beginning of the 1970's, when de Bergevin and Brunel \cite{Brunel_1972} demonstrated that x-rays are also sensitive to the distribution of electronic spins in magnetic materials by detecting antiferromagnetic Bragg reflections in NiO, thus confirming theoretical predictions by Platzman and Tzoar \cite{Platzman_1970}. Following these seminal findings, synchrotron-based x-ray magnetic scattering has been subsequently used as a powerful alternative to neutron scattering on several magnetically-ordered systems \cite{Brunel_1981,Gibbs_1985,Namikawa_1985,Gibbs_1986,Bohr_1987,Goldman_1987}, while around the same time the foundations for a comprehensive theory of resonant x-ray scattering (RXS) were laid by Blume and coworkers \cite{Blume_1985,Blume_1988,Hannon_1988}. Since its inception during the 1980's, resonant hard x-ray scattering has developed into an extremely versatile tool \cite{Materlik_1994,Carra_1994} for the selective study of ordering phenomena involving the charge, spin, orbital, and lattice degrees of freedom, often offering a unique perspective on their respective interplay.
\begin{marginnote}
\entry{RXS}{Resonant X-ray scattering}
\end{marginnote}

Historically, hard x-ray methods have anticipated soft x-ray methods due to the ${(\hbar \omega)}^{3}$ dependence of the x-ray attenuation length on photon energy $\hbar \omega$ ($\omega$ is the angular frequency of the radiation field), which eliminates the need for vacuum-based experimental chambers for energies $\hbar \omega \gtrsim 5$\,keV\footnote{For example, the attenuation length for x-rays propagating in atmosphere at 10\,keV (1\,keV) is $\sim\! 3$\,m (3\,mm) and, as a result, the loss in x-ray flux over 1 m (representing the typical dimension of a scattering diffractometer) is $\sim\!30$\% (100\%). Due to the continuous, power-law dependence of x-ray absorption on the photon energy, no precise cutoff can be defined for soft x-rays. However, conventional Cr-${K}_{\alpha}$ x-ray sources exist that operate in air at $\sim\! 5.4$\,keV, albeit requiring reduced source-to-sample and sample-to-detector distances. In addition, we note that hard x-rays beamlines still require vacuum conditions along the long pipes transporting the photons from the storage ring to the experimental chamber and from the chamber to the photon detectors.}.
\begin{figure}[t!]
\includegraphics[width=1\linewidth]{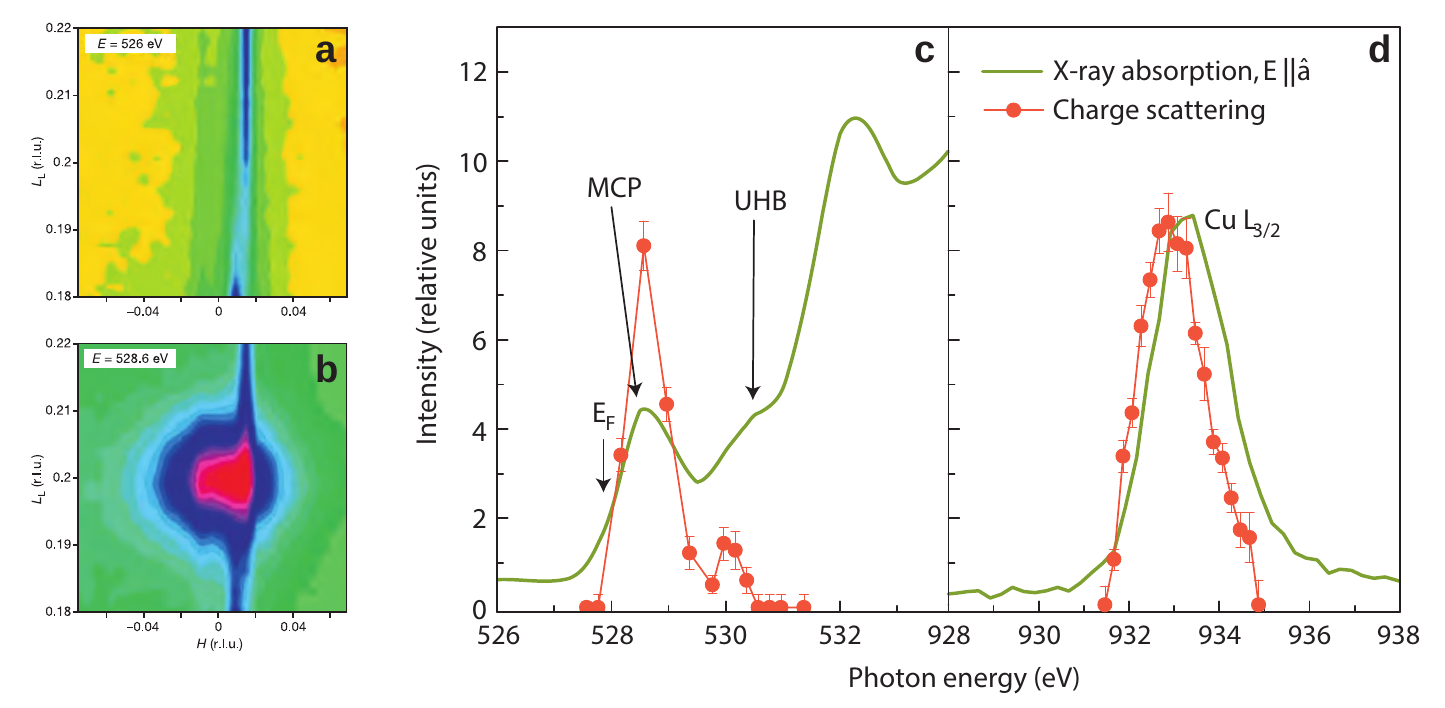}
\caption{\textbf{The first resonant soft x-ray scattering studies of charge order in cuprates.} (\textit{a,b}) Reciprocal space mapping of the $(H0L)$ plane in the non-superconducting cuprate compound Sr${}_{14}$Cu${}_{24}$O${}_{41}$ measured using soft x-ray scattering (\textit{a}) off-resonance (526\,eV) and (\textit{b}) at the main absorption peak for doped O-$2p$ holes (528.6\,eV, corresponding to the $1s \rightarrow 2p$ transition). At resonance (\textit{b}), a superlattice peak is revealed at $\mathbf{Q} \!=\! (0,0,0.2)$, signaling the crystallization of O-$2p$ holes into a static and periodic pattern (Wigner crystal). Readapted from Ref.\,\citen{abbamonte2004}. (\textit{c}) Resonant soft x-ray scattering measurements of the stripe order peak of La${}_{1.875}$Ba${}_{0.125}$CuO${}_{4}$ [located at $\mathbf{Q} \!\sim\! (0,25,0,1.5)$] across the O-$K$ edge ($1s \rightarrow 2p$, left), with highlighted the x-ray transitions into the mobile carrier states (mobile carrier peak, MCP) and upper Hubbard band (UHB). (\textit{d}) Same as in (\textit{c}), but at the Cu-${L}_{3}$ ($2p \rightarrow 3d$, right) absorption edge. The scattering peak intensity (red line and markers) is resonantly enhanced in the vicinity of the features of the absorption spectra (green line) corresponding to electronic transitions into the O and Cu sites in the CuO${}_{2}$ planes. Readapted from Ref.\,\citen{abbamonte2005}.}\label{RSXS_origin_fig}
\end{figure}
Soft x-ray absorption (XAS) was first pioneered using conventional x-ray sources and in a transmission geometry on thin films of rare earth metals and oxides, known for the large absorption cross-section and rich multiplet structure characterizing the ${M}_{4,5}$ edges ($3d \rightarrow 4f$ transitions) \cite{Mariot_1974,Bonnelle_1977}. Important advancements in XAS came with the use of synchrotron facilities \cite{Karnatak_1981}, which produced a higher x-ray flux and further enabled the control of linear and circular x-ray polarization, making it possible to detect magnetic x-ray dichroism effects \cite{vanderLaan_1986}. In the same years, improved detection schemes were being progressively adopted, such as partial electron yield \cite{Thole_1985}, total electron yield \cite{Kuiper_1988,deGroot_1989}, and fluorescence yield \cite{Troger_1990,Krol_1990}.

The successes of resonant hard x-ray scattering spurred the development of soft x-ray scattering instruments and methods, which presented clear benefits -- access to the electronic states and corresponding degrees of freedom controlling the electronic structure in transition metal and rare earth oxides, greater sensitivity to surface and interface effects, native control of incoming light polarization -- but also inherent complications -- limited accessible reciprocal space (and a consequent lower boundary of $\sim\! 3$\,\AA on the smallest measurable periodicity, at $\sim\! 2$\,keV), the incompatibility with crystal-based energy or polarization analyzers, the need for high vacuum environment. The rise of resonant soft x-ray scattering required special diffractometers capable of simultaneous, in-vacuum control of the sample and detector angles. The first concept of a two-circle diffractometer (with independent sample and detector rotations in the scattering plane) was pioneered already in the late 1980's \cite{Jark1988} and further advancements were realized throughout the 1990's and early 2000's at different facilities including Brookhaven National Lab (Ref.\,\citen{Kao1990} and later Ref.\,\citen{Andrivo_2006}), LURE (Ref.\,\citen{Tonnerre1995}), the  Synchrotron Radiation Center (Ref\,\citen{MacKay1996}), the European Synchrotron Radiation Facility (Ref.\,\citen{Durr1999}), the Daresbury Laboratory (Refs.\,\citen{Roper2001,Wilkins2003,Wilkins2003_LSMO}), BESSY (Ref.\,\citen{Grabis2003}), the Swiss Light Source (Ref.\,\citen{Staub2005}), and the Canadian Light Source (Ref.\,\citen{Hawthorn_revsciinst}). Motivated by the new insights produced by early resonant inelastic scattering studies at the Cu-$K$ edge ($\sim\! 8.9$\,keV) \cite{Hill1998,Abbamonte1999,Hasan2000,Kim2002}, the first resonant scattering measurements on cuprates in the soft x-ray regime were performed in 2002 by Abbamonte \textit{et al.} \cite{Abbamonte2002} on thin films of oxygen-doped La${}_{2}$CuO${}_{4}$, at the O-$K$ edge ($1s \rightarrow 2p $, $\hbar \omega \!\sim\! 530-550$\,eV) and Cu-${L}_{3,2}$ edge ($2p \rightarrow 3d $, $\hbar \omega \!\sim\! 920-960$\,eV). The sensitivity of these two absorption edges to the doped holes in cuprates had been demonstrated previously \cite{Nucker1988,deGroot_1989,Romberg1990,Chen_1992}, in particular through the identification of a mobile carrier peak (MCP) structure at $\hbar \omega \!\sim\! 528$\,eV, i.e. below the onset of the main O-$K$ edge, and representing  transitions onto the doped hole electronic states with large spectral weight on the O-$2p$ orbitals within the CuO${}_{2}$ planes. The study by Abbamonte \textit{et al.} explored momentum space in the ranges $(0,0,0.21) - (0,0,1.21)$ (orthogonal to the CuO${}_{2}$ planes) and $(0,0,0.6) - (0.3,0,0.6)$ (parallel to the CuO${}_{2}$ planes) and the evolution of the momentum-dependent interference fringes across the resonances was determined to be indicative of a rounding of the carrier density near the film-substrate interface. Most remarkably, the authors revealed the extent of resonant enhancement at these absorption edges in the cuprates, which amounts to a single doped hole in the CuO${}_{2}$ planes scattering as strongly as 82 electronic charges, with a resulting magnification of the experimental signal equal to the squared of the scattering amplitude, or ${82}^{2} > {10}^{3}$. This can be regarded as one of the key figures of merit for resonant x-ray methods, and represents the foundational mechanism underlying the success of RXS in detecting weak ordering phenomena (such as charge order) in the cuprates.

Later, Abbamonte \textit{et al.}\,\cite{abbamonte2004} used RXS to discover and characterize the ordering of doped holes in the non-superconducting cuprate compound Sr${}_{14}$Cu${}_{24}$O${}_{41}$. A peak in reciprocal space, representing the signature of a well-defined, periodic modulation of the electronic density along the reciprocal $L$ axis and with wavevector $\mathbf{Q} \!=\! (0,0,0.2)$, could only be measured at the O-$K$ pre-peak resonance at 528.6\,eV (Fig.\,\ref{RSXS_origin_fig}b) while no signal was detected once the photon energy was tuned off-resonance by only a few eV (Fig.\,\ref{RSXS_origin_fig}a). This work represents the very first RXS evidence of charge order in a copper-oxide compound, corresponding to the crystallization into a Wigner crystal phase, manifested as a modulation of the electronic charge triggered by electron-electron Coulomb interactions. This observation demonstrates the power of resonant scattering methods as opposed to conventional diffraction techniques, which, being non-resonant, are largely insensitive to subtle electronic ordering phenomena not involving the lattice degrees of freedom.

The first RXS study of charge order in one of the superconducting cuprate families came along in 2005, when Abbamonte \textit{et al.} \cite{abbamonte2005} investigated stripe order in LBCO. While the momentum structure of the stripe-ordered phases in this compound had been previously studied using neutron scattering \cite{Fujita_2002}, no direct information was yet available on the role and involvement of the electronic degrees of freedom, due to the fact that neutron scattering predominantly measures periodic distortions in the lattice. Abbamonte \textit{et al.} found a peak in reciprocal space at $(0.25,0,L)$, consistent with previous observations \cite{Fujita_2002} and with a weak dependence on the out-of-plane momentum $L$, thus confirming the prominent two-dimensional nature of the charge ordered state. Once again, this study reaffirmed the fundamental role played by the resonant enhancement: Figure\,\ref{RSXS_origin_fig}c and d show the intensity of the charge order peak (red markers) overlaid onto the absorption profiles (green lines) for the O-$K$ and Cu-${L}_{3}$ edges, respectively. The intensity for charge scattering is nonzero, i.e. detectable on top of the fluorescent background and above noise level, only near the Oxygen prepeak (MCP) and the Cu excitonic resonance. This work had profound implications for the understanding of charge order in cuprates, since it clarified that stripe order is predominantly configured as a modulation of the electronic density followed by a distortion of the lattice, which however represents a secondary effect, as demonstrated by the absence of any diffracted intensity away from the electronic resonances (in the soft x-ray range).
\begin{marginnote}
\entry{LBCO}{La${}_{2-x}$Ba${}_{x}$CuO${}_{4+\delta}$}
\end{marginnote}

In the following sections we will provide a brief description of the theory of resonant scattering and of the experimental scheme. For a more comprehensive treatise on these topics, we refer the reader to Refs.\,\citen{Blume_1985,Materlik_1994,Carra_1994,Fink_review,DiMatteo_2012} and references therein.

\subsection{Theory of resonant scattering}

Resonant x-ray scattering is a \textit{photon in -- photon out} technique, where photons get scattered from a material due to the interaction with the electronic clouds. For radiation-matter scattering to occur, the interaction Hamiltonian has to contain operator combinations of the kind $ {a}_{\nu}\left( \mathbf{q} \right) {a}_{\nu}^{\dagger} \left( \mathbf{q-Q} \right) $, where ${a}_{\nu}^{\dagger} \left( \mathbf{q} \right)$ [${a}_{\nu} \left( \mathbf{q} \right)$] is the operator creating [annihilating] a photon with wavevector $\mathbf{q}$, polarization state $\nu$, and frequency $\omega \!=\! c \left\vert \mathbf{q} \right\vert$. The effective nonrelativistic interaction Hamiltonian can be derived from the full electron-matter minimal coupling Hamiltonian, and reads: 
\begin{eqnarray}
{H}_{tot} &= {\sum}_{j} \left\lbrace {\frac{1}{2{m}_{e}}\left[{\mathbf{p}}_{j} - \frac{e}{c} \mathbf{A} ({\mathbf{r}}_{j}, t) \right]}^{2} + V ({\mathbf{r}}_{j}, t) \right\rbrace + {\sum}_{j \neq k} \frac{{e}^{2}}{{\vert{\mathbf{r}}_{j} - {\mathbf{r}}_{k}\vert}^{2}} + {H}_{\mathrm{EM}} \nonumber
\\
&= \underbrace{{H}_{\mathrm{el}} + {H}_{\mathrm{EM}}}_{{H}_{0}} + \underbrace{\frac{e}{{m}_{e} c} {\sum}_{j} \mathbf{A} ({\mathbf{r}}_{j}, t) \cdot {\mathbf{p}}_{j}}_{{H}_{\mathrm{int}}^{\mathrm{lin}}} + \underbrace{\frac{{e}^{2}}{2{m}_{e} {c}^{2}} {\sum}_{j} {{\mathbf{A}}^{2} ({\mathbf{r}}_{j}, t)}}_{{H}_{\mathrm{int}}^{\mathrm{quad}}}
\end{eqnarray}
where $e$ and $m$ are the fundamental electronic charge and mass, ${\mathbf{p}}_{j}$ and ${\mathbf{r}}_{j}$ represent the momentum and position coordinates of the \textit{j}-th electron respectively, and $V (\mathbf{r}, t)$ and ${e}^{2}/{\vert\mathbf{r} - \mathbf{r'}\vert}^{2}$ are the lattice potential and the Coulomb interaction terms, respectively. $\mathbf{A} (\mathbf{r}, t)$ represents the vector potential, ${H}_{\mathrm{el}} \!=\! {\sum}_{j} {\frac{1}{2{m}_{e}}}{\mathbf{p}}_{j}^{2} + {\sum}_{j} V ({\mathbf{r}}_{j}, t) + {\sum}_{j \neq k} \frac{{e}^{2}}{{\vert{\mathbf{r}}_{j} - {\mathbf{r}}_{k}\vert}^{2}} $ is the Hamiltonian of the electronic system, while ${H}_{\mathrm{EM}} \!=\! {\sum}_{\mathbf{q}, \nu} \hbar \omega \left[ {a}_{\nu}^{\dagger} \left( \mathbf{q} \right) {a}_{\nu} \left( \mathbf{q} \right) + 1/2 \right] $ is the Hamiltonian of the electromagnetic (EM) field alone. 

The interaction operators ${H}_{\mathrm{int}}^{\mathrm{lin}}$ and ${H}_{\mathrm{int}}^{\mathrm{quad}}$, which are respectively linear and quadratic in the vector potential, couple the electromagnetic field and the electronic degrees of freedom. At this point, we can use as basis set for the light-matter quantum system the states $ {| {\Psi}_{M} \rangle\!} \!=\! {| {\psi}_{m} \rangle\!}_{\mathrm{el}} \times {| {\phi}_{{\bar{n}}_{\mathbf{q},\nu}} \rangle\!}_{\mathrm{EM}} $, where ${| {\psi}_{m} \rangle\!}_{\mathrm{el}}$ represents the electronic part of the wavefunction (with eigenvalues ${\epsilon}_{m}$, and $m$ labeling a generic set of quantum numbers) while $ {| {\phi}_{{\bar{n}}_{\mathbf{q},\nu}} \rangle\!}_{\mathrm{EM}} $ indicates a photon state with photon occupation ${\bar{n}}_{\mathbf{q},\nu} \!=\! \left\lbrace {n}_{{\mathbf{q}}_{1}, {\nu}_{1}}, {n}_{{\mathbf{q}}_{2}, {\nu}_{2}}, {}_{\ldots} \right\rbrace $, corresponding to having $ {n}_{{\mathbf{q}}_{1}, {\nu}_{1}} $ photons with wavevector and polarization $\left( {\mathbf{q}}_{1}, {\nu}_{1} \right)$, $ {n}_{{\mathbf{q}}_{2}, {\nu}_{2}} $ photons with $\left( {\mathbf{q}}_{2}, {\nu}_{2} \right)$, and so on. In this notation, $M$ labels the global set of quantum numbers, i.e. $M = \left\lbrace m, \mathbf{q}, \nu \right\rbrace$. Note that due to the radiation-matter interaction, the states $ | {\Psi}_{M} \rangle$ are not eigenstates of the system, but they can be used as basis set in a perturbative scheme, in which case we define the unperturbed (i.e., with the interaction terms turned off) energy spectrum as ${E}_{M} \!=\! {\epsilon}_{m} + {\sum}_{\mathbf{q}, \nu} \left( {n}_{\mathbf{q}, \nu} \hbar {\omega}_{\mathbf{q}} + 1/2 \right) $. Within this framework, a scattering process is defined as a transition from an initial photon state ${| {\phi}_{i} \rangle\!}_{\mathrm{EM}}=\! | {}_{\ldots} \rangle | {n}_{{\mathbf{q}}_{\mathrm{in}},{\nu}_{\mathrm{in}}} \rangle | {n}_{{\mathbf{q}}_{\mathrm{out}},{\nu}_{\mathrm{out}}} \rangle | {}_{\ldots} \rangle $ to a final photon state ${| {\phi}_{f} \rangle\!}_{\mathrm{EM}}=\! | {}_{\ldots} \rangle | {n}_{{\mathbf{q}}_{\mathrm{in}},{\nu}_{\mathrm{in}}} - 1 \rangle | {n}_{{\mathbf{q}}_{\mathrm{out}},{\nu}_{\mathrm{out}}} + 1 \rangle | {}_{\ldots} \rangle $, corresponding to the annihilation of an incoming photon with wavevector and polarization $ \left( {\mathbf{q}}_{\mathrm{in}}, {\nu}_{\mathrm{in}} \right)$ and the concomitant creation of an outgoing photon $ \left( {\mathbf{q}}_{\mathrm{out}}, {\nu}_{\mathrm{out}} \right)$.
\begin{figure}[t!]
\includegraphics[width=0.9\linewidth]{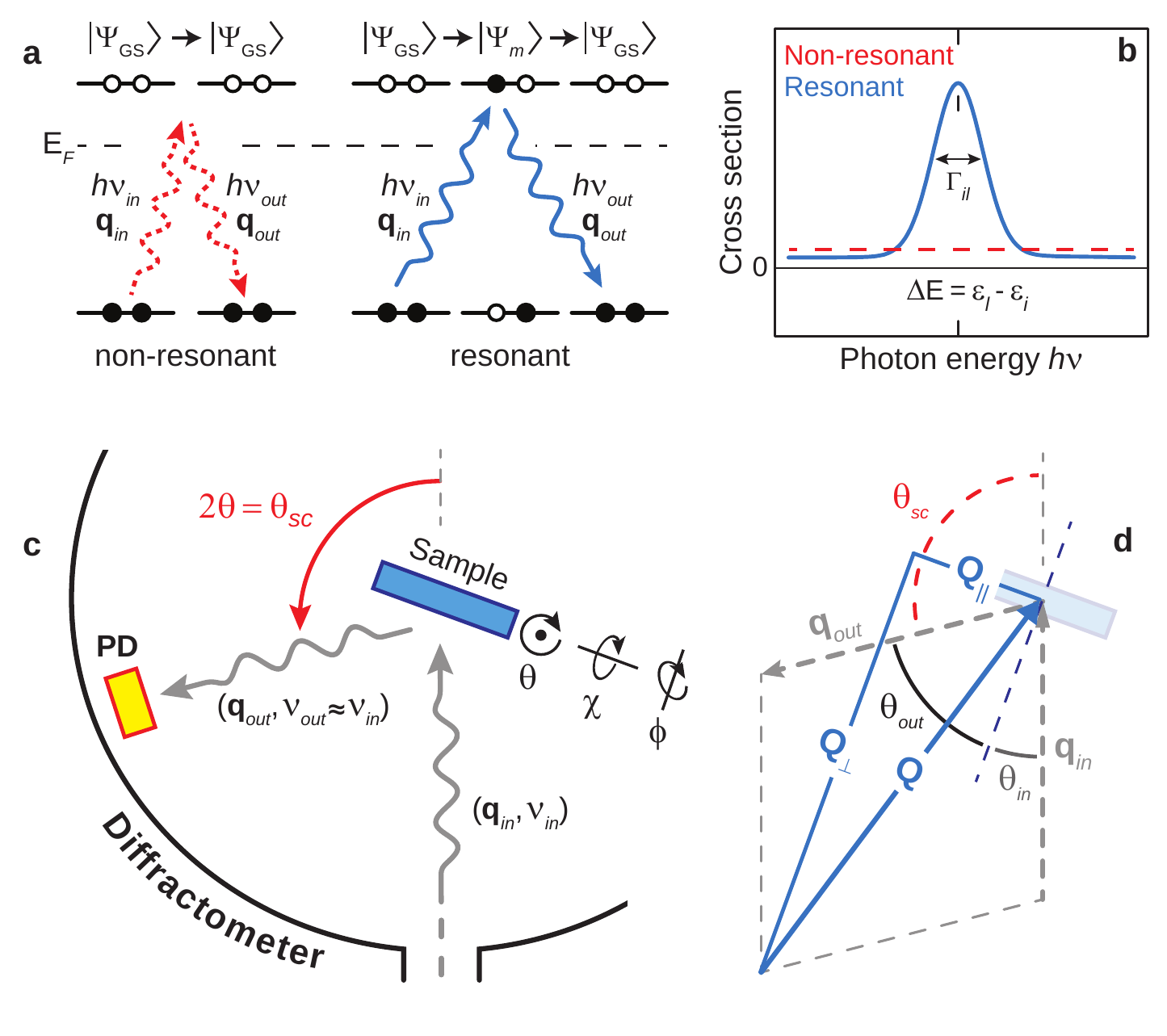}
\caption{\textbf{Resonant processes and scattering geometry in RXS.} (\textit{a}) In non-resonant scattering the excitation process does not involve intermediate states, while resonant scattering occurs whenever the incident photon energy is tuned to promote an electronic transition from the ground state ${\Psi}_{GS}$ to an intermediate state ${\Psi}_{m}$. The subsequent radiative recombination of the excited electron with the core hole results in the creation of an outgoing (scattered) photon. (\textit{b}) The different photon energy dependence of resonant and non-resonant processes, showing the enhancement occurring in the resonant channel near an electronic transition with energy $\Delta E$. (\textit{c}) Schematics of a conventional diffractometer, with kinematics for the scattering/diffraction process outlined in (\textit{d}).}\label{Mechanism_RSXS}
\end{figure}

Central in the theory of elastic scattering is the calculation of the probability of transition from a state $ {| {\Psi}_{i} \rangle\!} \!=\! {| {\psi}_{GS} \rangle\!}_{\mathrm{el}} \times {| {\phi}_{i} \rangle\!}_{\mathrm{EM}} $ to a state $ {| {\Psi}_{f} \rangle\!} \!=\! {| {\psi}_{GS} \rangle\!}_{\mathrm{el}} \times {| {\phi}_{f} \rangle\!}_{\mathrm{EM}} $, where the photon states ${| {\phi}_{i} \rangle\!}_{\mathrm{EM}}$ and ${| {\phi}_{f} \rangle\!}_{\mathrm{EM}}$ are as given in the previous paragraph, and where we further assume that the electronic part of the initial and final state is in the ground state ${| {\psi}_{GS} \rangle\!}_{\mathrm{el}}$. The transition probability ${w}_{i \rightarrow f}$ between the initial and final quantum states can be calculated using the generalized Fermi's golden rule \cite{Bruus_2011}:
\begin{equation}
{w}_{i \rightarrow f} = 2 \pi {\left\vert \langle {\Psi}_{i} | T | {\Psi}_{f} \rangle \right\vert}^{2} \delta \left( {E}_{f} - {E}_{i} \right)
\label{eq:FGR}
\end{equation}
where the $T$-matrix is defined as follows:
\begin{equation}
T = {H}_{\mathrm{int}} + {H}_{\mathrm{int}} \frac{1}{{E}_{i} - {H}_{0} + i \eta} {H}_{\mathrm{int}} + {H}_{\mathrm{int}} \frac{1}{{E}_{i} - {H}_{0} + i \eta} {H}_{\mathrm{int}} \frac{1}{{E}_{i} - {H}_{0} + i \eta} {H}_{\mathrm{int}} + {}_{\ldots}
\label{eq:T_matrix}
\end{equation}
Here the first operator on the right-hand side represents the first-order perturbation term, the second operator represents the second-order perturbation term, and so on. In Eq.\,\ref{eq:T_matrix}, ${H}_{\mathrm{int}} $ is the interaction operator and $ {H}_{0} $ is the unperturbed Hamiltonian -- in our case ${H}_{\mathrm{int}} \!=\! {H}_{\mathrm{int}}^{\mathrm{lin}} + {H}_{\mathrm{int}}^{\mathrm{quad}} $ and $ {H}_{0} \!=\! {H}_{\mathrm{el}} + {H}_{\mathrm{EM}} $, respectively. In scattering, we require operator combinations of the kind ${a}^{\dagger} a$ inside the $T$-matrix, which originate from interaction terms that are quadratic in the vector potential $\mathbf{A}$, since the latter can be expressed in second-quantized notation as $ \mathbf{A} (\mathbf{r}, t)\!\propto\!{\sum}_{\mathbf{q}, \nu} {\boldsymbol \varepsilon}_{\nu} \cdot \left[ \exp(i\mathbf{q}\cdot\mathbf{r} - i\omega t) \: {a}_{\nu}^{\dagger} \left( \mathbf{q} \right) +h.c. \right] $ (${\boldsymbol \varepsilon}_{\nu}$ denoting the polarization vector of the polarization state $\nu$). These combinations are generated by using the quadratic interaction operator ${H}_{\mathrm{int}}^{\mathrm{quad}}$ in the first-order term of Eq.\,\ref{eq:T_matrix}, and by the linear interaction operator ${H}_{\mathrm{int}}^{\mathrm{lin}}$ in the second-order term of Eq.\,\ref{eq:T_matrix}.
The corresponding first [${w}_{i \rightarrow f}^{(1)}$] and second [${w}_{i \rightarrow f}^{(2)}$] order perturbative transition probabilities can then be obtained from Eq.\,\ref{eq:FGR}:
\begin{eqnarray}
{w}_{i \rightarrow f}^{(1)} & = & 2 \pi {\left\vert \frac{{e}^{2}}{2{m}_{e} {c}^{2}} \langle {\Psi}_{i} | {\sum}_{j} {{\mathbf{A}}^{2} ({\mathbf{r}}_{j}, t)} | {\Psi}_{f} \rangle \right\vert}^{2} \label{eq:Wfi_nonres} \\
{w}_{i \rightarrow f}^{(2)} & = & 2 \pi {\left\vert {\left( \frac{e}{{m}_{e} c} \right)}^{2} {\sum}_{M} \frac{\langle {\Psi}_{i} \vert {\sum}_{j} \mathbf{A} ({\mathbf{r}}_{j}, t) \cdot {\mathbf{p}}_{j} \vert {\Psi}_{M} \rangle \langle {\Psi}_{M} \vert {\sum}_{k} \mathbf{A} ({\mathbf{r}}_{k}, t) \cdot {\mathbf{p}}_{k} \vert {\Psi}_{f} \rangle}{ {E}_{i} - {E}_{M} + i {\Gamma}_{M}} \right\vert}^{2} \label{eq:Wfi_res}
\end{eqnarray}
\noindent
where for convenience we have dropped the $\delta$-function (enforcing conservation of the total energy in the scattering process) originally present in Eq.\,\ref{eq:FGR}.

In Eq.\,\ref{eq:Wfi_res}, ${\Psi}_{M}$ represents a generic (excited) quantum state of the light-matter system, with corresponding energy ${E}_{M}$ and lifetime $\hbar/{\Gamma}_{M}$. The squared vector potential can then be expanded as:
\begin{eqnarray}
{{\mathbf{A}}^{2} (\mathbf{r}, t)} & \propto & {\sum}_{\mathbf{q}, \nu} {\boldsymbol \varepsilon}_{\nu} \left[ {e}^{ i (\mathbf{q} \cdot \mathbf{r} - \omega t)} \: {a}_{\nu}^{\dagger} \left( \mathbf{q} \right) +h.c. \right] \times {\sum}_{{\mathbf{q}}^{\prime}, {\nu}^{\prime}} {\boldsymbol \varepsilon}_{{\nu}^{\prime}} \left[ {e}^{ i ({\mathbf{q}}^{\prime} \cdot \mathbf{r} - {\omega}^{\prime} t)} \: {a}_{{\nu}^{\prime}}^{\dagger} \left( {\mathbf{q}}^{\prime} \right) +h.c. \right] \nonumber \\
 & \propto & \left( {\boldsymbol \varepsilon}_{{\nu}_{\mathrm{in}}}  \cdot {\boldsymbol \varepsilon}_{{\nu}_{\mathrm{out}}} \right) \left[ {e}^{ i ({\mathbf{q}}_{\mathrm{out}} \cdot \mathbf{r} - {\omega}_{\mathrm{out}} t)} \cdot {a}_{{\nu}_{\mathrm{out}}}^{\dagger} \left( {\mathbf{q}}_{\mathrm{out}} \right) \cdot {e}^{ -i ({\mathbf{q}}_{\mathrm{in}} \cdot \mathbf{r} - {\omega}_{\mathrm{in}} t)} \cdot {a}_{{\nu}_{\mathrm{in}}} \left( {\mathbf{q}}_{\mathrm{in}} \right) \right] \nonumber\\
 & \propto &\left( {\boldsymbol \varepsilon}_{{\nu}_{\mathrm{in}}}  \cdot {\boldsymbol \varepsilon}_{{\nu}_{\mathrm{out}}} \right) \times {e}^{\left[ i \left( {\mathbf{q}}_{\mathrm{out}} - {\mathbf{q}}_{\mathrm{in}} \right) \cdot \mathbf{r} \right] } \cdot {a}_{{\nu}_{\mathrm{out}}}^{\dagger} \left( {\mathbf{q}}_{\mathrm{out}} \right) {a}_{{\nu}_{\mathrm{in}}} \left( {\mathbf{q}}_{\mathrm{in}} \right)
\label{eq:A_squared}
\end{eqnarray}
where ${\boldsymbol \varepsilon}_{{\nu}_{\mathrm{in}}}$ and ${\boldsymbol \varepsilon}_{{\nu}_{\mathrm{out}}}$ represent the polarization vector of the incoming and scattered photons, and in the last step of Eq.\,\ref{eq:A_squared} we assumed the scattering process to be elastic, i.e. ${\omega}_{\mathrm{in}} \!=\! {\omega}_{\mathrm{out}}$.

At this point, by using the previous definitions for the initial and final states ${| {\Psi}_{i} \rangle\!}$ and ${| {\Psi}_{f} \rangle\!}$, together with the fact that ${E}_{i} \!=\! {\epsilon}_{\mathrm{GS}} + \left[ {n}_{{\mathbf{q}}_{\mathrm{in}}} \hbar {\omega}_{{\mathbf{q}}_{\mathrm{in}}} + 1/2 \right]$ and ${E}_{M} \!=\! {\epsilon}_{\mathrm{m}} + \left[ \left( {n}_{{\mathbf{q}}_{\mathrm{in}}} - 1 \right) \hbar {\omega}_{{\mathbf{q}}_{\mathrm{in}}} + 1/2 \right]$, and considering that $ {a}_{{\nu}_{\mathrm{out}}}^{\dagger} \left( {\mathbf{q}}_{\mathrm{out}} \right) {a}_{{\nu}_{\mathrm{in}}} \left( {\mathbf{q}}_{\mathrm{in}} \right) {| {\phi}_{i} \rangle\!}_{\mathrm{EM}} \propto {| {\phi}_{f} \rangle\!}_{\mathrm{EM}}$, we can rewrite Eqs.\,\ref{eq:Wfi_nonres} and \ref{eq:Wfi_res} as follows:
\begin{eqnarray}
{w}_{i \rightarrow f}^{(1)} & \propto & {\left\vert \langle {\psi}_{\mathrm{GS}} | {\sum}_{j} {e}^{-i \mathbf{Q} \cdot {\mathbf{r}}_{j}} | {\psi}_{\mathrm{GS}} \rangle \right\vert}^{2} \propto {\left\vert \langle {\psi}_{\mathrm{GS}} | \: \rho \left( \mathbf{Q} \right) | {\psi}_{\mathrm{GS}} \rangle \right\vert}^{2} \label{eq:Wfi_nonres_el} \\
{w}_{i \rightarrow f}^{(2)} & \propto & {\left\vert \sum\limits_{m} \sum\limits_{j,k} \frac{\langle {\psi}_{\mathrm{GS}} \vert {\boldsymbol \varepsilon}_{{\nu}_{\mathrm{in}}} \cdot {\mathbf{p}}_{j} \cdot {e}^{i {\mathbf{q}}_{\mathrm{in}} \cdot {\mathbf{r}}_{j}} \vert {\psi}_{m} \rangle \langle {\psi}_{m} \vert {\boldsymbol \varepsilon}_{{\nu}_{\mathrm{out}}} \cdot {\mathbf{p}}_{k} \cdot {e}^{-i {\mathbf{q}}_{\mathrm{out}} \cdot {\mathbf{r}}_{k}} \vert {\psi}_{\mathrm{GS}} \rangle}{ {\epsilon}_{\mathrm{GS}} - {\epsilon}_{m} + \hbar \omega + i {\Gamma}_{m}} \right\vert}^{2} \label{eq:Wfi_res_el}
\end{eqnarray}
where $\mathbf{Q} \!=\! {\mathbf{q}}_{\mathrm{in}} - {\mathbf{q}}_{\mathrm{out}}$ is the momentum transferred by the photon field to the sample, and $\rho \left( \mathbf{Q} \right)$ is the Fourier transform of the electron density operator $\rho \left( \mathbf{r} \right) \!=\! {\sum}_{j} \delta \left( \mathbf{r} - {\mathbf{r}}_{j} \right) $.

In the x-ray regime, the transition channel represented by Eq.\,\ref{eq:Wfi_res_el} involves the excitation of a high-energy many-body state with a core hole (${\psi}_{m}$). In a more intuitive perspective, this is equivalent to the excitation of a core electron into an intermediate state through absorption of the first photon, followed by re-emission of a (scattered) photon once the core hole is filled back. On the other hand, in the first-order perturbative term expressed by Eq.\,\ref{eq:Wfi_nonres}, the scattering process is instantaneous as it does not involve the excitation of an intermediate state. As a consequence, the first-order mechanism (known as Thomson scattering), proportional to the squared amplitude of the total electronic density in the ground state, is \textit{non-resonant} and controls the signal in conventional x-ray diffraction (XRD). The second-order process is instead \textit{resonant}, and is therefore associated to RXS.
\begin{marginnote}
\entry{XRD}{X-ray Diffraction}
\end{marginnote}

Equation\,\ref{eq:Wfi_res_el} is a particular version of the Kramers-Heisenberg formula, which represents the general solution to the problem of a photon scattering from an electron. It can be further simplified under the assumption that the x-ray excitation process is local, which implies that: (i) the local orbitals (at the lattice site $n$) can be used for the electronic basis set: ${\psi}_{m} (\mathbf{r}) \rightarrow {\chi}_{l}^{(n)} (\mathbf{r}) = {\chi}_{l} (\mathbf{r} - {\mathbf{R}}_{n}) $; (ii) that all matrix elements $ \langle {\chi}_{i}^{(m)} \vert \mathbf{p} \vert {\chi}_{l}^{(n)} \rangle \propto {\delta}_{m,n} $, i.e. that they vanish for orbitals belonging to different lattice sites; and (iii) that, due to the localization of the initial (core) electron around the position ${\mathbf{R}}_{n}$ of its parent atom, the phase due to the photon field can be approximated as ${e}^{i \mathbf{q} \cdot \mathbf{r}} \!\sim\! {e}^{i \mathbf{q} \cdot {\mathbf{R}}_{n}} $. At this stage, it is convenient to introduce a new quantity, the \textit{form factor} ${f}_{p q}$, which is a photon energy- and site-dependent complex tensor defined as follows:
\begin{equation}
{f}_{p q}^{(n)} \!\left( \hbar \omega \right) = \frac{{e}^{2}}{ {m}^{2} {c}^{2}} \: {\sum}_{i,l} \, \frac{ \langle {\chi}_{i}^{(n)} \vert {\mathrm{p}}_{q} \vert {\chi}_{l}^{(n)} \rangle \cdot \langle {\chi}_{l}^{(n)} \vert {\mathrm{p}}_{p} \vert {\chi}_{i}^{(n)} \rangle}{\hbar \omega - ({\epsilon}_{l}^{(n)}-{\epsilon}_{i}^{(n)})+i{\Gamma}_{il}} .
\label{form_factor}
\end{equation}
\noindent
Here ${\chi}_{i}^{(n)}$ and ${\chi}_{l}^{(n)}$ represent the initial and intermediate single-particle electronic states at site ${\mathbf{R}}_{n}$ (with energies ${\epsilon}_{i}^{(n)}$ and ${\epsilon}_{l}^{(n)}$, respectively) involved in the light-induced transition $i \rightarrow l$. ${\Gamma}_{il}$ is the inverse lifetime ($\hbar / {\tau}_{il}$) of the intermediate state with an electron in ${\chi}_{l}^{(n)}$ and a hole in ${\chi}_{i}^{(n)}$. 

\noindent
The resonant scattering cross-section, through the form factor ${f}_{pq}^{(n)} \left( \hbar \omega \right)$, can be shown to bear a close connection to the x-ray absorption (XAS), which is a first order process in the radiation-matter interaction Hamiltonian:
\begin{eqnarray}
{I}^{\mathrm{XAS}} \left( \hbar \omega \right) \!& \propto &\! - \frac{1}{{\omega}^{2}} \times \mathrm{Im} \left[ {\sum}_{n} {\sum}_{p} \: {\left( {\boldsymbol \varepsilon}_{{\nu}_{\mathrm{in}}} \right)}_{p} \cdot {f}_{pp}^{(n)} \left( \hbar \omega \right) \right]
\label{XAS_def}
\\
{I}^{\mathrm{RXS}} \left( \mathbf{Q}, \hbar \omega \right) \!& \propto &\! {\left\vert {\sum}_{p q} {\left( {\boldsymbol \varepsilon}_{{\nu}_{\mathrm{in}}} \right)}_{p} \cdot \left[ {\sum}_{n} \: {f}_{pq}^{(n)} \left( \hbar \omega \right) {e}^{i \mathbf{Q} {\textstyle\cdot} \: {\mathbf{R}}_{n}} \right] \cdot {\left( {\boldsymbol \varepsilon}_{{\nu}_{\mathrm{out}}} \right)}_{q} \right\vert}^{2}\nonumber\\
\!& = &\! {\left\vert {\sum}_{p q} {\left( {\boldsymbol \varepsilon}_{{\nu}_{\mathrm{in}}} \right)}_{p} \cdot {F}_{pq} \left( \hbar \omega \right) \cdot {\left( {\boldsymbol \varepsilon}_{{\nu}_{\mathrm{out}}} \right)}_{q} \right\vert}^{2} \: .
\label{RSXS_def}
\end{eqnarray}
In Eq.\,\ref{RSXS_def}, ${F}_{pq}$ represents the scattering tensor, which \textit{is not} a local quantity (does not depend on the lattice position ${\mathbf{R}}_{n}$) and is more directly related to the physical observable in RXS experiments (${I}^{\mathrm{RXS}}$). Moreover, from the above equations it follows that XAS only depends on the incoming light polarization ${\boldsymbol \varepsilon}_{{\nu}_{\mathrm{in}}} $, whereas the RXS signal depends on the outgoing light polarization ${\boldsymbol \varepsilon}_{{\nu}_{\mathrm{out}}} $, as well.

The difference between resonant and non-resonant scattering is schematized in Fig.\,\ref{Mechanism_RSXS}(a). The mechanism corresponding to XRD involves a single step, in virtue of its first-order nature; conversely RXS, being a second-order transition, proceeds in two stages involving an intermediate state. This clearly reflects in the very different photon energy ($h\nu$) dependence of the two channels [see Fig.\,\ref{Mechanism_RSXS}(b)]: whereas XRD is nearly energy-independent (red dashed curve), the cross-section for RXS is strongly peaked around the energy of the electronic transition (blue curve), where the experimental signal undergoes a strong enhancement (while decaying to zero away from the resonance). This usually occurs in correspondence of an absorption edge, i.e. when electronic transition from a deeply bound core state into the valence band (and beyond into the continuum) take place. As a consequence, RXS gains a strong sensitivity [as large as 82-fold in the cuprates \cite{Abbamonte2002,abbamonte2005}] to partial modulations of the charge density involving a single electronic band, whereas the XRD signal reflects the total electronic density (see Eq.\,\ref{eq:Wfi_nonres_el}) and therefore suffers from a weak sensitivity to spatial variations of the latter, unless they are accompanied by a distortion of the lattice, which would involve all the electrons (core and valence).

\subsection{The experimental scheme}

In an actual scattering or diffraction experiment, a monochromatic x-ray beam with wavevector ${\mathbf{q}}_{\mathrm{in}}$, photon energy $\hbar {\omega}_{\mathrm{in}}\!=\!c \cdot {q}_{\mathrm{in}} $ and polarization $ {\boldsymbol \varepsilon}_{{\nu}_{\mathrm{in}}} $ impinges on a sample and a scattered photon will be detected along the direction of the wavevector ${\mathbf{q}}_{\mathrm{out}}$ using an energy-integrating photon detector (PD) or an energy-resolving spectrometer (see Fig.\,\ref{Mechanism_RSXS}c). At the end of the process, a net momentum and energy have been transferred to the sample, which can be derived from the corresponding conservation laws:
\begin{eqnarray}
{h\nu}_{\mathrm{in}} \!&=&\! {h\nu}_{\mathrm{out}} + \Delta E
\label{Comin:eq:totenegy_RSXS}
\\
{\mathbf{q}}_{\mathrm{in}} \!&=&\! {\mathbf{q}}_{\mathrm{out}} + \mathbf{Q}.
\label{Comin:eq:totmomentum_RSXS}
\end{eqnarray}
\noindent
In the case of elastic scattering ${h\nu}_{\mathrm{in}}\!=\!{h\nu}_{\mathrm{out}}$ and there is no energy transfer with the sample ($\Delta E\!=\!0$), whereas the case $\Delta E\!\neq\!0$ defines inelastic scattering events. Strictly speaking, elastic scattering probes the static component of the charge and magnetization density occurring in the system under study, whereas inelastic scattering is sensitive to dynamical processes and low-energy excitations. However, due to the spectrometer-characteristic finite energy resolution $\delta E$ (which in the soft x-ray regime ranges between 30\,meV and 1\,eV, while hard x-ray spectrometers reach down to 1\,meV), purely elastic scattering cannot be experimentally accessed, and it is more appropriate to use the term quasi-elastic scattering, which probes a regime which is static up to a timescale $\tau \sim \hbar / \delta E$. From a more practical standpoint, the energy-integrated measurement in most cases yields a reliable representation of the momentum structure of the ordered state, due to fact that the inelastic part of the spectra usually evolves very smoothly and can be discarded as background in RXS, especially if it exhibits a different temperature dependence with respect to the zero energy-loss feature (see also discussion of Fig.\,\ref{RSXS_YBCO_Ghiringhelli_fig}).

Hereafter, we will focus on the momentum structure of RXS measurements and, unless otherwise specified, will assume the use of energy-integrated mode. From Eq.\,\ref{Comin:eq:totmomentum_RSXS}, and using ${h\nu}_{\mathrm{in}}\!=\!{h\nu}_{\mathrm{out}}$, the magnitude of the exchanged momentum can be expressed as $Q\!=\!2\,{q}_{\mathrm{in}} \times \sin \left( {\theta}_{\mathrm{sc}} / 2 \right) $ [${\theta}_{\mathrm{sc}}$ is the scattering angle, see Fig.\,\ref{Mechanism_RSXS}(c)]. Projecting $\mathbf{Q}$ into the plane defining the sample surface then yields the in-plane (${\mathbf{Q}}_{\parallel}$) and out-of-plane (${\mathbf{Q}}_{\perp}$) components of the transferred momentum, which will be often referenced in the rest of this work [see Fig.\,\ref{Mechanism_RSXS}(d)]. The realization of this geometry is illustrated in Fig.\,\ref{Mechanism_RSXS}(c), and described in greater detail in Ref.\,\citen{Hawthorn_revsciinst}. In general the photon detector moves on a single-circle, that is, a single angular goniometer (the corresponding variable is often denominated $2 \theta$ and corresponds to the scattering angle ${\theta}_{\mathrm{sc}}$). The sample stage usually involves translational motion (xyz) and various rotations, whose number defines the type of diffractometer. Two-circle diffractometers represent the most common choice for soft x-ray experiments, featuring a single angular motion (with angular variable denominated $\theta$) for the sample (${1}^{\mathrm{st}}$ circle), whose axis is perpendicular to the scattering plane (the one spanned by the vectors ${\mathbf{q}}_{\mathrm{in}}$ and ${\mathbf{q}}_{\mathrm{out}}$), in addition to the detector rotation ($2 \theta$), which represents the ${2}^{\mathrm{nd}}$ circle. The first generation of soft x-ray diffractometers also includes two additional rotational degrees of freedom (denominated $\chi$ and $\phi$, see again Fig.\,\ref{Mechanism_RSXS}c), allowing a fine alignment of the sample axes with respect to the scattering plane (however, $\chi$ and $\phi$ typically cover a limited angular range of ${-5}^{\circ}$ to ${5}^{\circ}$). New designs are being developed using different geometries to extend the angular range and control for the sample orientation. 
Diffractometers with more circles (up to 6 -- 3 for the sample, 3 for the detector) covering an ample angular range are routinely used at higher photon energies (hard x-rays, $\hbar \omega \!>\!7$\,keV), where the added complication of the all-vacuum environment required for soft x-rays is lifted. Consequently, hard x-ray diffractometers can typically access a wider portion of reciprocal space.

Typically, the experimental signal is comprised of both resonant and non-resonant contributions, and in the two possible regimes $ w_{fi}^{(XRD)} \!\gg\! w_{fi}^{(RXS)} $ or $ w_{fi}^{(RXS)} \!\gg\! w_{fi}^{(XRD)} $ one ends up probing different phenomena (see again Fig.\,\ref{Mechanism_RSXS}b). In the first case, where non-resonant processes are dominant, all electrons contribute equally to the measured signal, which is therefore simply proportional to the atomic number $Z$. The diffraction signal will be dominated by the core electrons, which usually outnumber the valence ones ($ {n}_{\mathrm{core}} \gg {n}_{\mathrm{valence}} $), with the exception of lighter elements  which as a result are not probed very effectively in XRD. In addition, since core states are very tightly bound to their parent nucleus, in a conventional diffraction experiment one mainly probes the ionic lattice in reciprocal space, which is why XRD is used primarily for structural studies.
In the second case, the scattering process has a strong enhancement in correspondence of a very specific electronic transition. As a result the signal bears the signature of the electronic distribution (in reciprocal space) of the final state of such transition. This characteristic of resonant scattering allows it to be not only element-specific (whenever the absorption edges of different chemical species are spaced sufficiently apart in photon energy), but also orbital-selective. This unique capability of RXS has been established and employed in many different systems. Charge-ordering in cuprates \cite{abbamonte2005} and cobaltates \cite{Tjeng_2005}, and orbital-ordering in the manganites \cite{Wilkins2003_LSMO,Dhesi2004,Thomas_2004}, are among the most spectacular case studies.

\section{CHARGE ORDER IN CUPRATES -- A NEVER-ENDING JOURNEY}

\subsection{A resurging phenomenology: charge-density-waves in YBCO}

Fifteen years after the original discovery of stripe order by Tranquada \textit{et al.}, charge order had been observed in the doping region around 12\% hole doping directly in the 3 families of La-, Bi- and Ca-based cuprates: LNSCO \cite{tranquada1995,Tranquada_1996,Tranquada_1997,vZimmermann_1998,Wilkins_2011}, LBCO \cite{Fujita_2002,abbamonte2005,Wilkins_2011,Hucker_2011}, and LESCO \cite{Fink2009,Fink2011} using neutron and x-ray scattering (in the case of Refs\,\citen{Fink2009,Fink2011} resonant x-ray scattering was used, in particular); Bi2212 \cite{hoffman2002,McElroy2003_PhysicaC,Hoffman2003_PhysicaC,howald2003,howald2003_PRB,vershinin2004,hanaguri2004,McElroy2005_PRL,koshaka2007,parker2010,daSilvaNeto,daSilvaNeto_PhysicaC,Fujita_2012},  Na-CCOC \cite{hanaguri2004,koshaka2007}, and Bi2201 \cite{wise2008} using STM. However, no clear evidence of charge order had been found in the YBCO compounds, where the possibility of introducing doped carriers via the fractional filling of the Cu-O chain layer helps reducing the electronic inhomogeneity in the CuO${}_{2}$ planes \cite{Liang_2000}.
\begin{marginnote}
\entry{LESCO}{La${}_{2-x-y}$Eu${}_{y}$Sr${}_{x}$Cu${}_{2}$O${}_{4+\delta}$}
\end{marginnote}
\begin{figure}[t!]
\includegraphics[width=1\linewidth]{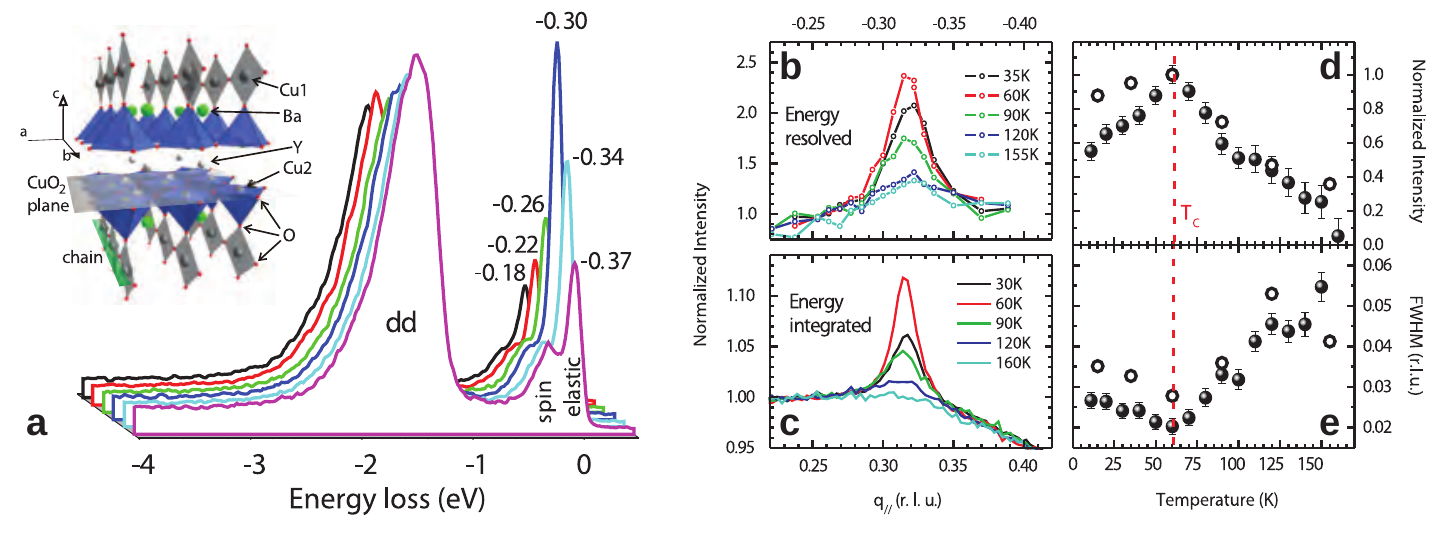}
\caption{\textbf{Resonant x-ray scattering discovery of charge-density-waves in (Nd,Y)Ba${}_{2}$Cu${}_{3}$O${}_{6+x}$.} (\textit{a}) Resonant inelastic x-ray scattering (RIXS) measurements of underdoped Nd${}_{1.2}$Ba${}_{1.8}$Cu${}_{3}$O${}_{7}$ (${T}_{\mathrm{c}} \!=\! 65$\,K), as a function of energy loss and momentum [along $(H00)$], showing the emergence of a quasi-elastic peak around $H \!\sim\! -0.31$\,r.l.u.; inset: three-dimensional view of the lattice structure of YBa${}_{2}$Cu${}_{3}$O${}_{6+x}$. (\textit{b,c}) Momentum dependence of the quasi-elastic (\textit{b}) and of the energy-integrated RXS intensity (\textit{c}) for a series of temperatures across the superconducting transition ${T}_{\mathrm{c}} $. (\textit{d,e}) Temperature evolution of the charge order peak intensity (\textit{d}) and full-width-at-half-maximum (\textit{e}), showing a cusp at ${T}_{\mathrm{c}} $, thereby providing evidence of competition between charge order and superconductivity.
Readapted from Ref.\,\citen{ghiringhelli2012}.}\label{RSXS_YBCO_Ghiringhelli_fig}
\end{figure}

The first indication of electronic order in YBCO came in 2007 thanks to a series of pioneering measurements at high magnetic fields (up to 62\,T) and low temperatures (down to 1.5\,K) revealing the presence of quantum oscillations in the Hall resistance ${R}_{xy}$ in the normal state of underdoped YBa${}_{2}$Cu${}_{3}$O${}_{6.51}$ ($p \!=\! 0.10$, ${T}_{\mathrm{c}} \!=\! 57$\,K) samples \cite{doiron2007,leboeuf2007}. The frequency of these oscillations vs. the inverse magnetic field provided evidence for the emergence of small Fermi pockets at high fields \cite{doiron2007}, arguably of electron-like nature in light of the negative sign for the Hall coefficient ${R}_{H}$ \cite{leboeuf2007}. These findings indicated a change in the Fermi surface topology from the large hole-like Fermi `barrels' in the overdoped regime \cite{Hussey_2003,Plate_2005} to smaller pockets in the underdoped regime, thereby suggesting \textit{``a reconstruction of the Fermi surface caused by the onset of a density-wave phase, as is thought to occur in the electron-doped copper oxides near the onset of antiferromagnetic order"} \cite{leboeuf2007}. The striking similarity between the Hall and Seebeck coefficients of YBCO and those of LESCO \cite{Taillefer_2009,chang2010,laliberte2011} made a strong case for a form of order in YBCO akin to the stripe order in LESCO. The first direct evidence that the Fermi surface reconstruction is due to charge order was provided by high-field nuclear magnetic resonance (NMR) measurements on YBa${}_{2}$Cu${}_{3}$O${}_{6.54}$ ($p \!=\! 0.104$, ${T}_{\mathrm{c}} \!=\! 57$\,K), where a splitting in the ${}^{63}$Cu(2) lines was observed at 28.5\,T and below 50\,K, signaling a change in the quadrupole frequency which was ascribed to a periodic variation in the charge density at planar Cu sites or at the oxygen sites bridging them \cite{wu2011}. The simplest scenario seemingly compatible with the NMR results was a unidirectional density modulation with $4a$ (four unit cells) periodicity. Furthermore, the observation of the charge-order-induced NMR splitting for strong fields perpendicular but not parallel to the conducting CuO${}_{2}$ planes provided evidence of a competition between superconductivity and charge order.

The quantum oscillations and NMR results seemed to point to the necessity of completely suppressing superconductivity to observe the emergence of a seemingly competing charge ordered state. Following early RXS explorations by Hawthorn \textit{et al.}\,\cite{Hawthorn2011}, the first direct observation of charge density wave in reciprocal space, and at zero magnetic field, was obtained in 2012 by Ghiringhelli \textit{et al.} \cite{ghiringhelli2012} using energy-resolved and energy-integrated RXS on a series of YBCO doping levels around 12\%. The central experimental data uncovering the momentum location, and therefore the periodicity, of charge order in YBCO are shown in Fig.\,\ref{RSXS_YBCO_Ghiringhelli_fig}a, and consist of a series of scans of the x-ray energy loss (horizontal axis) for different values of the planar projection ($H$) of the momentum $ \mathbf{Q} \!=\! (H,K,L)$ along the $(H0L)$ direction in reciprocal space. The energy structure of the resonant inelastic x-ray scattering spectra bears three main contributions: (i) a quasi-elastic line at zero energy loss; (ii) a low-energy peak/shoulder representing spin excitations; and (iii) intra-band (dd) particle-hole excitations. While the spectral weight of the spin and dd excitations is nearly momentum-independent, a clear enhancement of the quasi-elastic line can be seen around a planar momentum $ H \!\sim\! -0.31$ r.l.u., which reveals the presence of a periodic modulation of the electronic density -- representing the momentum structure of a zero-field ordered state possibly connected to the one previously identified with NMR [recent high-field diffraction measurements, which will be discussed later, helped clarify the nature of this connection \cite{Gerber2015}]. The RXS data further revealed charge modulations to be present along both planar crystallographic axes, albeit with different amplitudes (as discussed more extensively in Ref.\,\citen{Blanco2014}). The same momentum-space peak could be equally well identified following the quasi-elastic component in energy resolved spectra (Fig.\,\ref{RSXS_YBCO_Ghiringhelli_fig}b) and the energy-integrated RXS (Fig.\,\ref{RSXS_YBCO_Ghiringhelli_fig}c), which confirms: (i) the importance of the resonant enhancement at the Cu-${L}_{3}$ edge; (ii) only the quasi-elastic scattering contributes to a pronounced momentum-resolved  structure, as previously discussed.

The detailed temperature dependence of energy-resolved and energy-integrated RXS momentum scans in Figs.\,\ref{RSXS_YBCO_Ghiringhelli_fig}b and c points to an onset temperature $T \!\sim\! 150$\,K and, most importantly, to a partial suppression of the charge order peak below the superconducting transition temperature ${T}_{\mathrm{c}}$. This finding provides direct evidence of a competition between superconductivity and charge order, and is further substantiated by the temperature evolution of the charge order peak intensity and full-width-at-half-maximum (FWHM, inversely proportional to the correlation lengths) reported in Fig.\,\ref{RSXS_YBCO_Ghiringhelli_fig}d and e for the case of energy-resolved and energy-integrated RXS, respectively. The intensity and FWHM data indicate a clear weakening of the charge ordered state both in its amplitude and spatial correlations at the emergence of the superconducting state. Around the same time, the wavevector of the charge-density-wave in YBCO, and its competition with the superconducting state were measured by non-resonant hard x-ray diffraction experiments, which revealed an enhancement of the charge order amplitude below ${T}_{\mathrm{c}}$ when superconductivity was actively weakened by applying magnetic fields up to 17\,T \cite{chang2012}. Field-induced enhancement of charge order was similarly reported in LBCO, using hard x-ray diffraction \cite{Hucker_2013}.

The discovery and identification of the charge order state in YBCO represented a breakthrough in the field of copper-oxide superconductors, as it suggested that the charge ordered state could be a defining instability of the CuO${}_{2}$ planes, and strongly contributed to revitalizing this research direction. Stimulated by the possibility to provide a unifying view of charge order in the cuprates, intense efforts at the experimental and theoretical level were put forth in order to identify a common thread across the different manifestations of charge order that emerged over the years.
\begin{figure}[t!]
\includegraphics[width=1\linewidth]{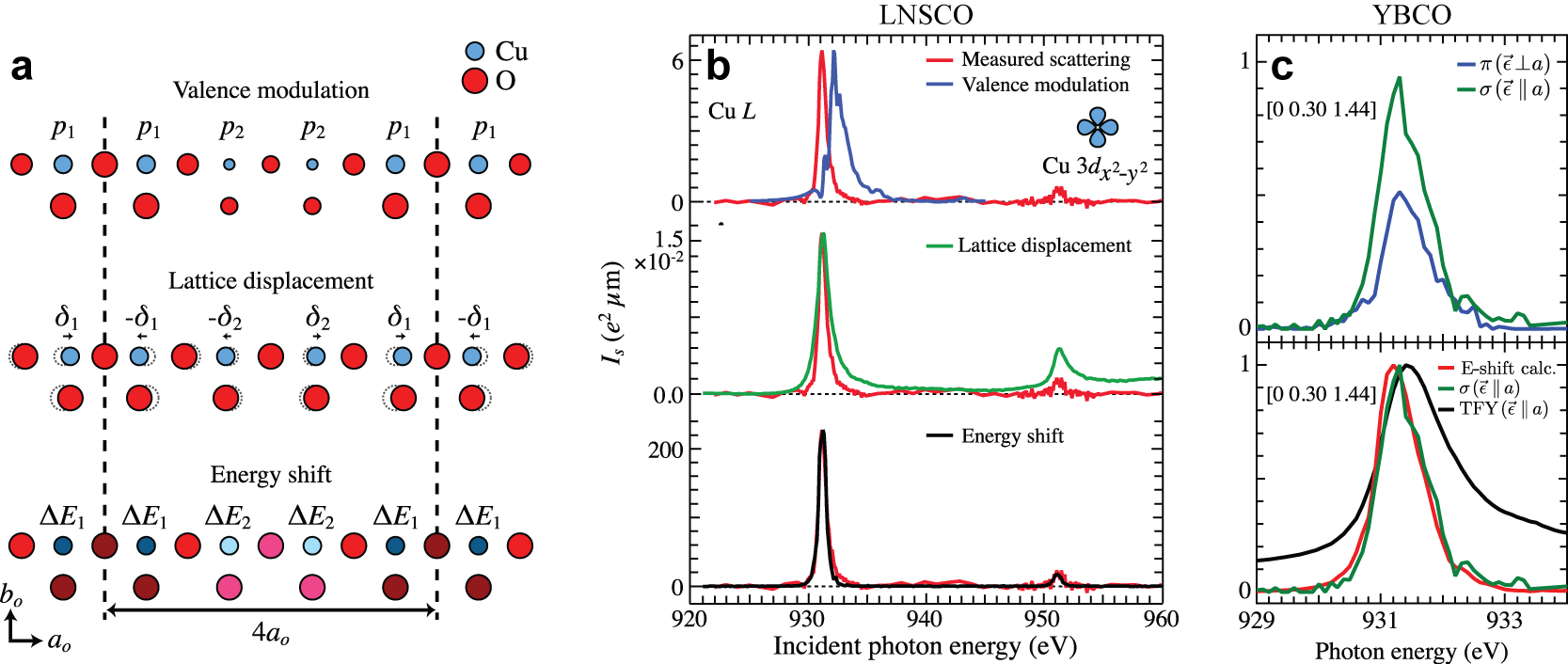}
\caption{\textbf{Photon energy-dependent resonant scattering from charge order in cuprates}. (\textit{a}) Schematic representation of the possible microscopic contributions to the RXS signal: valence modulation (top); lattice displacement (middle); transition energy shift (bottom). (\textit{b}) Comparison of the photon energy-dependent RXS intensity from the stripe order peak in LBCO to calculations from the models introduced in (\textit{a}), showing that the resonant scattering signal is predominantly controlled by spatial variations in the x-ray transition energies. Readapted from Ref.\,\citen{achkar2013}. 
(\textit{c}) Photon energy-dependent RXS signal from charge order in YBCO for horizontal ($\pi$) and vertical ($\sigma$) incoming polarization (top) and comparison between experiment and the prediction of the energy-shift model  for polarization along the \textbf{a} axis. Readapted from Ref.\,\citen{achkar2012}.}\label{RSXS_E_shifts_fig}
\end{figure}
A series of studies analyzed the photon energy dependence of the charge order signal from LESCO \cite{Fink2009}, LNSCO \cite{achkar2013}, and YBCO \cite{achkar2012}, which is inherited from the photon energy dependence of the complex form factor $f \left( \hbar \omega \right)$ (see Eqs.\,\ref{form_factor} and \ref{RSXS_def}). The imaginary part of the latter is directly related to the x-ray absorption signal, and the real part can be extracted by a Kramers-Kronig transformation. For a transition into a single Cu-${3d}_{{x}^{2}-{y}^{2}}$ hole (which is the case at the Cu-${L}_{3}$ edge in the cuprates) there is no multiplet structure, and the (site-dependent) form factor can be approximated by a simple Lorentzian lineshape as ${f}^{(n)} \!\sim\! {A}_{n} {\left( \hbar \omega - {\varepsilon}_{n} + i \Gamma \right)}^{-1} $, so that the RXS intensity can be expressed as:
\begin{equation}
{I}^{\mathrm{RXS}} \left( \mathbf{Q}, \hbar \omega  \right) \propto {\left\vert {\sum}_{n} \: {f}^{(n)} \left( \hbar \omega  \right) {e}^{i \mathbf{Q} {\textstyle\cdot} \: {\mathbf{R}}_{n}} \right\vert}^{2} = {\left\vert {\sum}_{n} \: \frac{{A}_{n}}{\left( \hbar \omega  - {\varepsilon}_{n} + i \Gamma \right)} {e}^{i \mathbf{Q} {\textstyle\cdot} \: {\mathbf{R}}_{n}} \right\vert}^{2}
\label{RSXS_def_Achkar}
\end{equation}
where the polarization dependence has been neglected for simplicity. Eq.\,\ref{RSXS_def_Achkar} shows that there are three variables controlling the RXS signal which can vary spatially: (i) the transition amplitude ${A}_{n}$; (ii) the lattice position ${\mathbf{R}}_{n}$; and (iii) the transition energy ${\varepsilon}_{n}$. Correspondingly, Achkar \textit{et al.} evaluated the impact on the RXS signal of assuming a periodic modulation for (see Fig.\,\ref{RSXS_E_shifts_fig}a): (i) the valence modulation; (ii) the lattice displacements; (iii) the transition energy shifts (this term was introduced in Refs\,\citen{achkar2012,achkar2013}). The comparison between the experimental RXS peak amplitude and the model calculations vs. photon energy is shown in Fig.\,\ref{RSXS_E_shifts_fig}b for LNSCO and in Fig.\,\ref{RSXS_E_shifts_fig}c for YBCO. In both cases, it was determined that the RXS signal is predominantly controlled by the spatial variation of the x-ray transition energy shifts, while in the earlier work on LESCO the RXS lineshape was explained with the nonlinear increase of the charge-carrier peak as a function of doping $p$, due to a reduction of correlation effects ($U$) at higher doping concentrations. More recently, an alternative theoretical framework has been devised that explains the photon energy-dependent RXS signal by accounting for the delocalized character of intermediate states \cite{Benjamin2013}.

A closely following series of studies provided a complete mapping of the detailed characteristics of this phenomenology -- the relative amplitude of the order parameter between YBCO and LBCO \cite{Thampy_2013}, the connection between charge order and magnetic instabilities \cite{Blackburn2013,Blanco2013,Hucker_2014}, the origin of finite correlations and the role of disorder in the chain layer \cite{Achkar_quenched_2013}, the feedback on lattice dynamics \cite{Bakr_2013,LeTacon2013,Blackburn2013_IXS} -- as a function of doping, temperature, and magnetic field. This tremendous amount of progress made it possible to shed new light on the role of charge order within the phase diagram and its connection to coexisting and neighbouring electronic orders and phases. The complete doping dependence of the salient properties of charge order in YBCO, and its comparison to La-based cuprates, were reported by Blanco-Canosa \textit{et al.}\,\cite{Blanco2014} and Hucker \textit{et al.}\,\cite{Hucker_2014} .

One of the open questions is the connection between the charge order seen in high magnetic fields via NMR \cite{wu2011,wu2013} and the charge modulations seen in zero field via x-ray diffraction. A field-induced thermodynamic phase transition was detected in the sound velocity of underdoped YBCO \cite{leboeuf2013}, suggesting that the high-field and low-field states are distinct, and in particular that the high-field charge order is two-dimensional in nature. NMR measurements, owing to their ability to probe both the high-field and the zero-field regimes, also established that charge order in these two regimes manifests itself as distinct phases, which furthermore coexist at low temperatures and high fields \cite{wu2015}. A recent x-ray diffraction study in pulsed magnetic fields up to 28 T revealed that there is indeed a field-induced crossover from a short-ranged order to a long-ranged charge modulations with different period along the \textbf{c} axis \cite{Gerber2015}.

\begin{marginnote}
\entry{NdBCO}{NdBa${}_{2}$Cu${}_{3}$O${}_{6+x}$}
\end{marginnote}

%
%

\subsection{Unifying real and reciprocal space: Bi2201 and Bi2212}

The discovery of CDWs in YBCO reinforced the idea that charge order might be a genuinely universal instability of the CuO${}_{2}$ planes, and the combination of real-space (STM) and reciprocal-space (neutron scattering, XRD, RXS) techniques had provided support for one and the same underlying general phenomenology in underdoped cuprates around 12\,\% hole doping. Despite such mounting evidence and the several unifying traits linking the various cuprate families, the real-space phenomenology of charge order in Bi-based compounds appeared very granular, at variance with the well-defined structures in reciprocal space as observed by scattering probes in La- and Y-based cuprates. The very different probing depth -- few Angstroms for STM vs. hundreds of nanometer (and more) for scattering techniques -- and a possible dichotomy between surface and bulk [as observed for charge-density waves along the Cu-O bond direction in LSCO \cite{Wu2012} and along the zone diagonal in Bi2201 \cite{Rosen_Comin}], also contributed to a perceived disconnect between the domains (real space/surface vs. momentum space/bulk) and materials (Bi- vs. La- and Y-based cuprates) investigated by these two classes of experimental methods.

\begin{figure}[t!]
\includegraphics[width=1\linewidth]{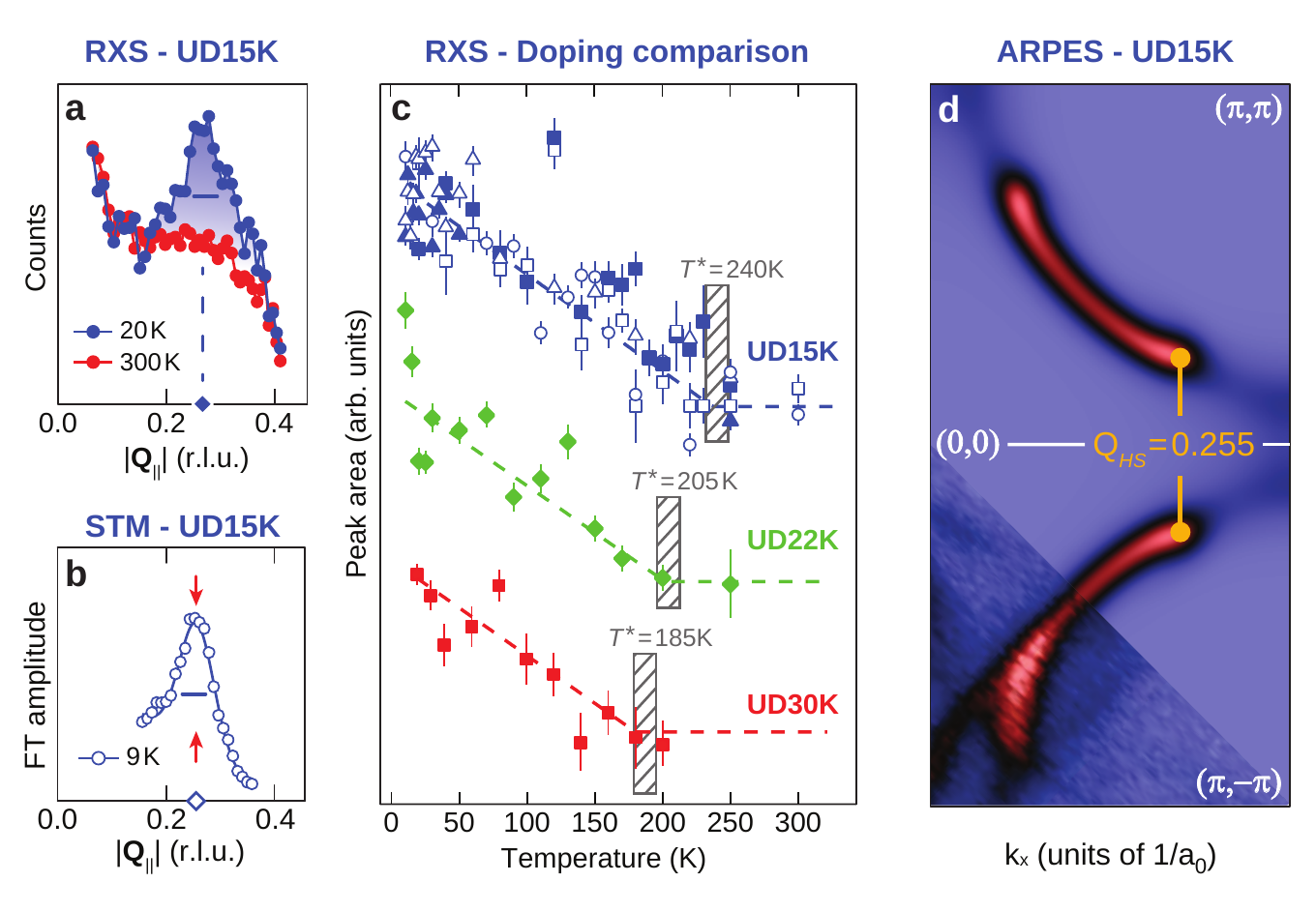}
\caption{\textbf{RXS and STM joint evidence of charge order in Bi${}_{2}$Sr${}_{2}$CuO${}_{6+\delta}$.} (\textit{a}) RXS scans along $(H00)$ for an underdoped (${T}_{\mathrm{c}} \!\sim\! 15$\,K) Bi2201 at low (blue) and high (red) temperature, showing the emergence of a broad charge order peak around $H \!\sim\! 0.25$ r.l.u.. (\textit{b}) $(H00)$ linecut of the Fourier-transformed differential tunnelling conductance on freshly-cleaved surfaces of the same Bi2201 samples, indicating the equivalence of the charge modulations detected using STM. (\textit{c}) Temperature dependence of the charge order peak intensity (from RXS) for three doping levels (${T}_{\mathrm{c}} \!\sim\! 15$\,K, $22$\,K, and $30$\,K) in the under-to-optimal doping range, and correlation of the charge order onset with the pseudogap temperature ${T}^{*}$ (grey boxes). (\textit{d}) Experimental (bottom) and calculated (top) Fermi surface for UD15K Bi2201, highlighting the Fermi arcs characterizing the pseudogap regime, and the wavevector (yellow connector) linking the arc-tips, or `hot-spots', in agreement with the RXS experimental results. Readapted from Ref.\,\citen{Comin_Science}.}\label{RSXS_Bi2201_fig}
\end{figure}

Two recent works \cite{Comin_Science,dSN_Science} addressed this aspect by investigating charge order on the same materials using STM and RXS. Due to the complications in exposing atomically flat surfaces in any cuprate material other than Bi-based compounds, these two studies were performed on La-doped Bi2201 (in the doping range $0.11 \!<\! p \!<\! 0.14$) \cite{Comin_Science} and Bi2212 (in the doping range $0.07 \!<\! p \!<\! 0.13$) \cite{dSN_Science}. In Bi2201, charge order was found with a wavevector between 0.243 and 0.265 r.l.u. for decreasing doping, whereas in Bi2212 the ordering wavevector spans across a more extended range, from 0.25 to 0.31 r.l.u., a finding which was later corroborated by the detection of charge order also in optimally-doped Bi2212 ($p \!\sim\! 0.16$, ${T}_{\mathrm{c}} \!=\! 98$\,K) \cite{Hashimoto_2014}. Most importantly, both studies demonstrated that the LDOS modulations imaged by STM and the reflections measured by RXS in reciprocal space originate from the very same microscopic entity. This is shown in Fig.\,\ref{RSXS_Bi2201_fig}a and b by the comparison between the RXS momentum scans (panel a) in underdoped Bi2201 (${T}{\mathrm{c}} = 15$\,K, $p \!\sim\! 0.11$), revealing a charge order peak emerging at low temperature at $\mathbf{Q} \!\sim\! (0.265,0,L)$, and the Fourier-transformed differential conductance data (panel b) taken on the very same samples, which exhibit a peak at the same momentum value. A similar correspondence was found in Bi2212. The temperature dependence of the RXS charge order peak for three different Bi2201 doping levels is reported in Fig.\,\ref{RSXS_Bi2201_fig}c, showing a very gradual onset of the density modulations which occurs on a temperature scale which is proximate to the pseudogap temperature ${T}^{*}$ as determined from Knight shift measurements \cite{Kawasaki2010}. In Bi2212, RXS data also reveal a rather slow temperature evolution of the peak intensity, which is accompanied by a drop below ${T}_{\mathrm{c}}$. Although this drop is smaller than its analogue in YBCO \cite{ghiringhelli2012,chang2012,achkar2012}, the competition with superconductivity is clearly demonstrated by temperature-dependent STM measurements that also reveal that the charge order is more pronounced for the unoccupied states \cite{dSN_Science}. Furthermore, in Bi2201 comparison between the RXS results and the Fermi surface measured using angle-resolved photoemission spectroscopy (ARPES) revealed a quantitative link between the observed charge order wavevector and the momentum vector connecting the tips of the Fermi arcs, which in this case coincide with the so-called `hot-spots' (see Fig.\,\ref{RSXS_Bi2201_fig}d). This correspondence suggests a possible link between the density modulations and the low-energy electronic structure, an element which is consistent with the doping evolution of the wavevector and which had been hinted by previous studies \cite{shen2005,wise2008}. Recent theoretical works have discussed the possible special role played by the hot-spots in the context of a magnetically-driven charge order instability \cite{Efetov_NatPhys,Sachdev2013,Davis_Lee_2013,Efetov_PRB,Chubukov2014}, as well as the possibility that charge order might arise from a $2 Q$ instability of the antinodal points within a pair-density-wave framework \cite{Lee2014}.
Additionally, we note that the RXS detection of charge order around $Q \!\sim\! 0.25$\,r.l.u. in Bi2201 confirms previously unpublished findings of a phonon anomaly at the same momentum location using inelastic x-ray scattering \cite{Bonnoit2012}, which might be suggestive of a related anomaly in the electronic susceptibility or otherwise of a very strongly momentum-dependent electron-phonon coupling mechanism at work.

\subsection{Towards a universal phenomenology: charge order in Hg1201 and NCCO}
\begin{figure}[t!]
\includegraphics[width=0.9\linewidth]{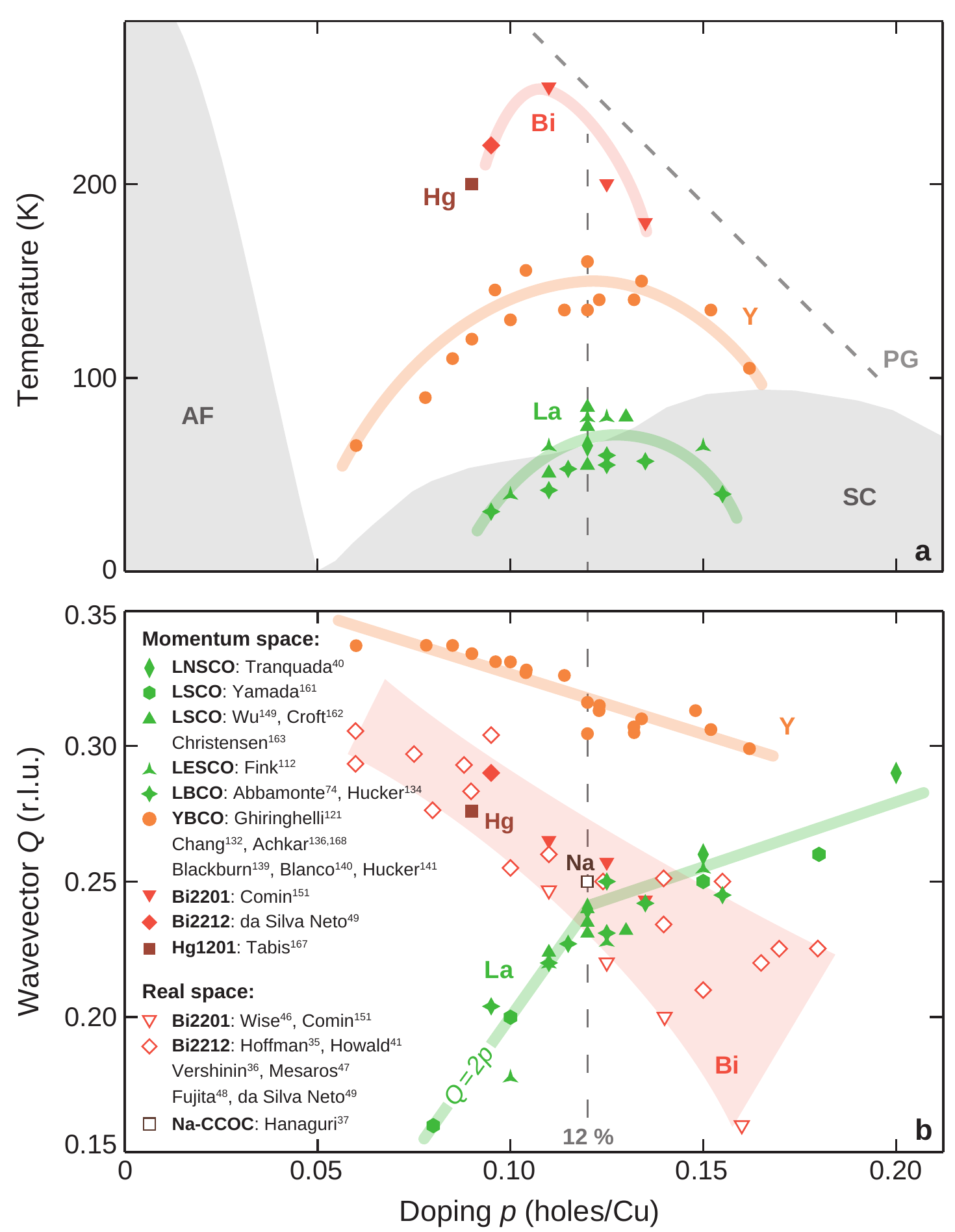}
\caption{\textbf{Charge order onset temperature and wavevector for all cuprate families.} (\textit{a}) Onset temperatures of charge order vs. hole doping, for all cuprate families. The shaded graphics in the background outline the boundaries of the antiferromagnetic (AF), superconducting (SC), and pseudogap (PG) regimes in YBCO. (\textit{b}) Doping dependence of the charge order wavevector (the LNSCO and LSCO experimental points from Refs.\,\citen{Tranquada_1997,Yamada_1998} are calculated from the position of the spin ordering wavevector). Full symbols are from momentum-resolved probes (RXS, XRD, neutron scattering), while open symbols are from real-space methods (STM). Colored lines are guides-to-the-eye; the vertical dashed line marks the location of the doping $p \!=\! 0.12$.}\label{CDW_all}
\end{figure}

In the wake of the results on Y- and Bi-based cuprates, evidence had been mounting that charge order could be a universal phenomenon in the cuprates. In 2012, Wu \textit{et al.} detected charge order in the prototypical superconductor LSCO, around a doping of 12\,\% \cite{Wu2012}, a discovery which revealed that the low-temperature tetragonal structure in La-based cuprates is not essential for the appearance of stripe order (but perhaps relevant for stabilizing it). In this study, the comparison between resonant and non-resonant scattering data was interpreted in terms of a surface enhancement of the stripe order in LSCO. More recently, stripe order was found in the bulk of LSCO using hard x-ray diffraction \cite{Croft_2014,Christensen_2014} and resonant x-ray scattering \cite{Thampy_2014}.
\begin{marginnote}
\entry{Hg1201}{HgBa${}_{2}$CuO${}_{4+\delta}$}
\entry{NCCO}{Nd${}_{2-x}$Ce${}_{x}$CuO${}_{4+\delta}$}
\end{marginnote}

Having exhausted all other families, Hg-based cuprates remained the last hole-doped family to be investigated. In particular, HgBa${}_{2}$CuO${}_{4+\delta}$ (Hg1201) represents the only series of compounds with a pristine, tetragonal unit cell, therefore possessing the highest structural symmetry among all cuprates. The first indirect evidence of broken translational symmetry in Hg1201 came from the discovery of Fermi-surface reconstruction via high-field measurements of the Hall and Seebeck coefficients in underdoped Hg1201 \cite{Doiron2013}, followed by the detection of quantum oscillations \cite{Barisic_2013}. In 2014, Tabis \textit{et al.} reported the very first evidence of charge order below 200\,K in underdoped ($p \!\sim\! 0.09$, ${T}_{\mathrm{c}} \!=\! 72$\,K) Hg1201 from RXS and RIXS measurements \cite{Tabis2014}. The charge order peak was found at $Q \!\sim\! 0.28 $\,r.l.u. --  comparable to the values found in Bi2212 and intermediate between YBCO and La-based compounds -- and required using resonant excitation, indicating that charge order is a rather subtle phenomenon in the Hg1201. This instance, combined with the fact that Hg1201 exhibits record-high ${T}_{\mathrm{c}}$ among single-layered cuprates, suggests a possible anticorrelation between charge order and superconductivity. Furthermore, the study revealed a direct connection between the onset of charge correlations and ${T}^{**}$, \textit{``the temperature at which the Seebeck coefficient reaches its maximum value, and below which conventional Fermi-liquid planar charge transport is observed"}. It also suggested a possible common origin of the main instabilities in the CuO${}_{2}$ planes, namely \textit{``the possibility that the sequence of ordering tendencies ($q \!=\! 0$ order precedes charge order, which in turn precedes superconducting order) and the phase diagram as a whole are driven by short-range antiferromagnetic correlations"}. Lastly, Tabis \textit{et al.} successfully established the link between the charge order-induced reconstruction of the Fermi surface and the size of the electron pockets observed by quantum oscillations in YBCO \cite{doiron2007,leboeuf2007} and Hg1201 \cite{Barisic_2013}, and were able to correctly predict the QO frequencies over an extended doping range.

Figure \ref{CDW_all} provides a graphical compendium of all charge order observations in the hole-doped cuprates for what concerns the doping dependence of the onset temperature (Fig.\,\ref{CDW_all}a) and of the wavevector (Fig.\,\ref{CDW_all}b) as observed using both spatial- (open symbols) and momentum-resolved (full symbols) probes. The shaded phase diagram in Fig.\,\ref{CDW_all}a is representative of YBCO and includes the antiferromagnetic (AF) phase near zero doping and the superconducting (SC) dome at higher hole doping levels. All coloured lines are guides-to-the-eye, except for the $Q \!=\! 2p$ line which interpolates the doping dependence of the ordering wavevector in LSCO below 12\,\% doping according to the picture of perfect stripes with 1/2 hole per unit cell along the charged rivers. Note that the points from Refs.\,\citen{Tranquada_1997,Yamada_1998} are from neutron scattering measurements of the antiferromagnetic peaks in LSCO, from which the charge order wavevector ${Q}_{\mathrm{CO}}$ was derived using the phenomenological relation ${Q}_{\mathrm{CO}} \!=\! 2 \, {Q}_{\mathrm{AF}}$. It is clear from the phase diagram of Fig.\,\ref{CDW_all}a how in all cases charge order emerges at the highest temperature scales around 12\,\% hole doping and spans a doping range tentatively bound by two endpoints\footnote{Note however that recently charge order has been observed in YBCO in the very underdoped regime, at $p \!\sim\! 0.058$ (${T}_{\mathrm{c}} \!\sim\! 12.6$\,K), along the $K$ reciprocal axis only, with wavevector ${Q}_{\mathrm{CO}} \!\sim\! 0.337$\,r.l.u. and onset temperature ${T}_{\mathrm{CO}} \!\sim\! 65$\,K \cite{Achkar_2015}.} \cite{Sebastian_review,He2014,Fujita2014_QCP,Ramshaw2015}, thereby hinting at a possible intimate connection between this phenomenology and quantum criticality \cite{Chubukov_1994,Sachdev2000}. Interestingly, the onset temperatures appear to scale inversely with the charge order amplitudes and correspondingly correlation lengths (which are maximum for La-based cuprates, intermediate for YBCO, and weak in the case of Bi-based compounds). At the same time, the extent of the suppression of ${T}_{\mathrm{c}}$ near 12\,\% doping is larger for the families exhibiting stronger charge order, consistent with the experimentally-observed competition between these two instabilities. Finally, we note that two main phenomenological differences exist between the stripy La-based cuprates and all other families: (i) in La-based compounds charge order and incommensurate antiferromagnetic spin order are simultaneously present, whereas they are mutually exclusive in the other compounds; (ii) the doping evolution of the charge order wavevector exhibits the opposite sign for La-cuprates with respect to the other families, which are putatively compatible with a nesting scenario (see also guides-to-the-eye in Fig.\,\ref{CDW_all}b). There are therefore strong indications for a common charge instability in all families of doped cuprates, but at the same time there are also stark differences between the manifestations of charge order in the conventional stripy compounds and in all other families.

With charge order consistently detected across all flavours of hole-doped cuprates around 12\,\% doping, the next frontier to be crossed was represented by the exploration of the electron-doped side of the phase diagram. Early insights came from inelastic x-ray scattering work \cite{Dastuto_2002} reporting anomalies in the dispersion of optical phonons in Nd${}_{2-x}$Ce${}_{x}$CuO${}_{4+\delta}$ (NCCO), followed by quantum oscillation studies providing evidence for a reconstructed small-pocket Fermi surface in the same compound \cite{Helm2009} and, more recently, by evidence from time-resolved studies of a broken-symmetry phase \cite{Hinton2013_NCCO}. Recently, the first hints of charge order were reported by Lee \textit{et al.} \cite{Lee2014_RIXS} and Ishii \textit{et al.} \cite{Ishii_2014}, who used RIXS to map out the momentum-dependent structure of low-energy bosonic excitations in NCCO, for electron-doping values spanning the phase diagram from the antiferromagnetic to the superconducting phase. In both studies, RIXS measurements revealed the presence of spin waves associated to the antiferromagnetic order at low doping, which evolved into paramagnon excitations in the superconducting state ($x \!\sim\! 0.15$). Around this doping, an additional branch of a rapidly-dispersing, gapped excitation was found, which was interpreted as a particle-hole excitations in Ref.\,\citen{Ishii_2014} and as a charge amplitude mode in Ref.\,\citen{Lee2014_RIXS}, therefore suggesting the possible signature of a symmetry-broken quantum state over an extended temperature range beyond the superconducting phase. 

\begin{figure}[t!]
\includegraphics[width=0.75\linewidth]{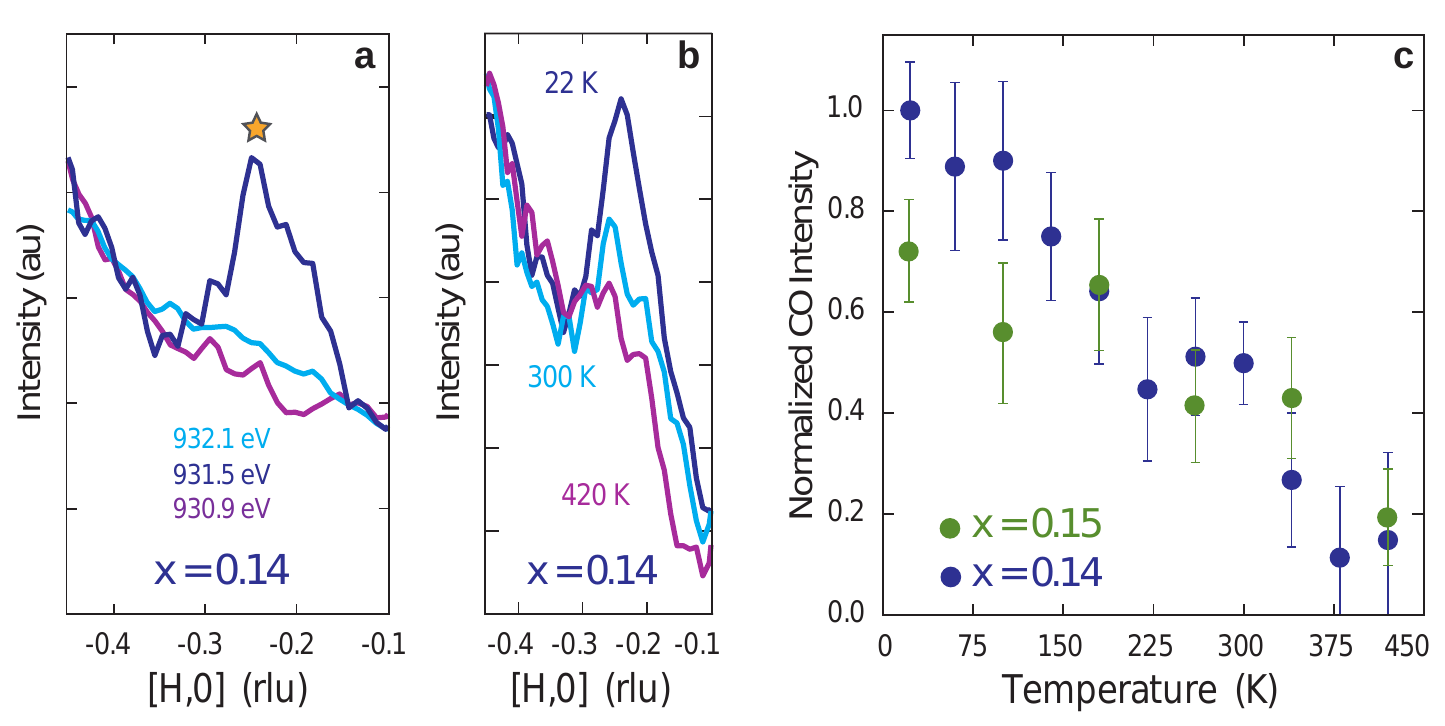}
\caption{\textbf{Resonant scattering study of charge order in the electron-doped cuprate NCCO.} (\textit{a,b}) Energy- (\textit{a}) and temperature-dependent (\textit{b}) RXS scans [along $(H00)$] of the charge order peak around $H \!\sim\! 0.25$ r.l.u. in superconducting Nd${}_{2-x}$Ce${}_{x}$CuO${}_{4+\delta}$ with $x \!=\! 0.14$. (\textit{c}) Gradual temperature evolution of the charge order intensity, with a tentative onset temperature around 350-400\,K. Readapted from Ref.\,\citen{dSN_NCCO_Science}.}\label{RSXS_NCCO_fig}
\end{figure}

Direct evidence for charge order in NCCO was obtained by Da Silva Neto \textit{et al.} \cite{dSN_NCCO_Science} by means of  RXS measurements in reciprocal space, which revealed a broad reflection at a wavevector $Q \!\sim\! 0.25$\,r.l.u., i.e. very proximate to the findings in hole-doped cuprates, once again pointing to a unified phenomenology and a possible universal instability of the CuO${}_{2}$ planes. The RXS scans for a superconducting NCCO sample ($x \!=\! 0.14$) are shown in Fig.\,\ref{RSXS_NCCO_fig}a and b as a function of photon energy and temperature, respectively. The charge order peaks appear to be rather broad, with a correlation length $\xi \!\sim\! 25-35$\,\AA, and a very smooth evolution as a function of temperature and a finite peak amplitude seemingly persisting up to rather high temperatures (350-400\,K), as reported in Fig.\,\ref{RSXS_NCCO_fig}c. Such a high onset temperature rules out a direct connection of charge order with the pseudogap phenomenology in electron-doped cuprates; instead, the charge order signal appears to partly correlate with the temperature evolution of the antiferromagnetic fluctuations in the same material \cite{Motoyama_2007}, uncovering a possible connection between the charge and magnetic instabilities, as previously suggested for the hole-doped cuprates \cite{Metlitski2010,Sachdev2013,Tabis2014,Efetov_NatPhys,Efetov_PRB}. Most importantly, and independently of the detailed temperature dependence, the detection of charge order in NCCO conclusively demonstrates that this phenomenon is truly universal in the cuprates, establishing a powerful and robust paradigm for the physics of the lightly-doped CuO${}_{2}$ planes, which transcends the asymmetry inherent to the different orbital character of doped holes (largely on O-$2p$ states) vs. doped electrons (occupying the Cu-derived upper Hubbard band).

\subsection{What hides behind a peak: charge order patterns and symmetries}

As the case for the universality of charge order was receiving increasingly supporting evidence, the theoretical understanding of the origin of this phenomenon and of its interplay with coexisting instabilities, as well as the detailed exploration of the microscopic structure of the ordered state, all regained a central role in the context of the physics of copper-based high-temperature superconductors. \cite{Efetov_NatPhys,Sachdev2013,Davis_Lee_2013,Efetov_PRB,Levin_2013,Bulut_2013,Kivelson_2013,DallaTorre_NJP,Lee2014,Chubukov2014,Norman2014}

In a single-band model charge order would be a scalar field, which can be expressed as the periodic modulation of the occupation of a given electronic orbital as a function of the spatial coordinate. However, the low-energy electronic structure of the CuO${}_{2}$ involves three different orbitals: Cu-${3d}_{{x}^{2} - {y}^{2}}$ (at the Copper sites), O-${2p}_{x}$ (at the horizontally-bridging Oxygen sites), and O-${2p}_{y}$ (at the vertically-bridging Oxygen sites). Consequently, charge order can be expressed as a vector field with three components \cite{Metlitski2010,Sachdev2013}: (i) a site-centered modulation, corresponding to an extra charge residing on the Cu-$3d$ orbital (Fig.\,\ref{RSXS_symmetry_YBCO_fig}c, top); (ii) an extended \textit{s'}-wave bond-order, where the spatially-modulated density is on the O-$2p$ states, and the maxima along the x and y directions coincide (Fig.\,\ref{RSXS_symmetry_YBCO_fig}c, middle); (iii) a \textit{d}-wave bond-order, where the charge modulation changes sign between x- and y-coordinated oxygen atoms, and the maxima are shifted by a half wavelength (Fig.\,\ref{RSXS_symmetry_YBCO_fig}c, bottom). The notation, as well as the denomination of `symmetry terms' for the components introduced above, follow from the resulting angular distribution of the phases within the CuO${}_{4}$ plaquette; this is encoded via an internal momentum variable $\mathbf{k}$ (restricted to the first Brillouin zone) in addition to the lattice momentum $\mathbf{Q}$, in the definition of the charge order parameter: $\Delta_{\mathrm{CDW}} (\mathbf{k},\mathbf{Q}) \!=\! \left\langle {c}^{\dagger}_{\mathbf{k+Q/2}} \cdot {c}_{\mathbf{k-Q/2}} \right \rangle $. This definition enables a full decomposition of the orbital-dependent modulation of the electronic density as a linear combination of the symmetry components introduced above, which therefore serve as a basis set for the charge order parameter, so that the latter can be expressed as \cite{Sachdev2013}:
\begin{figure}[t!]
\includegraphics[width=1\linewidth]{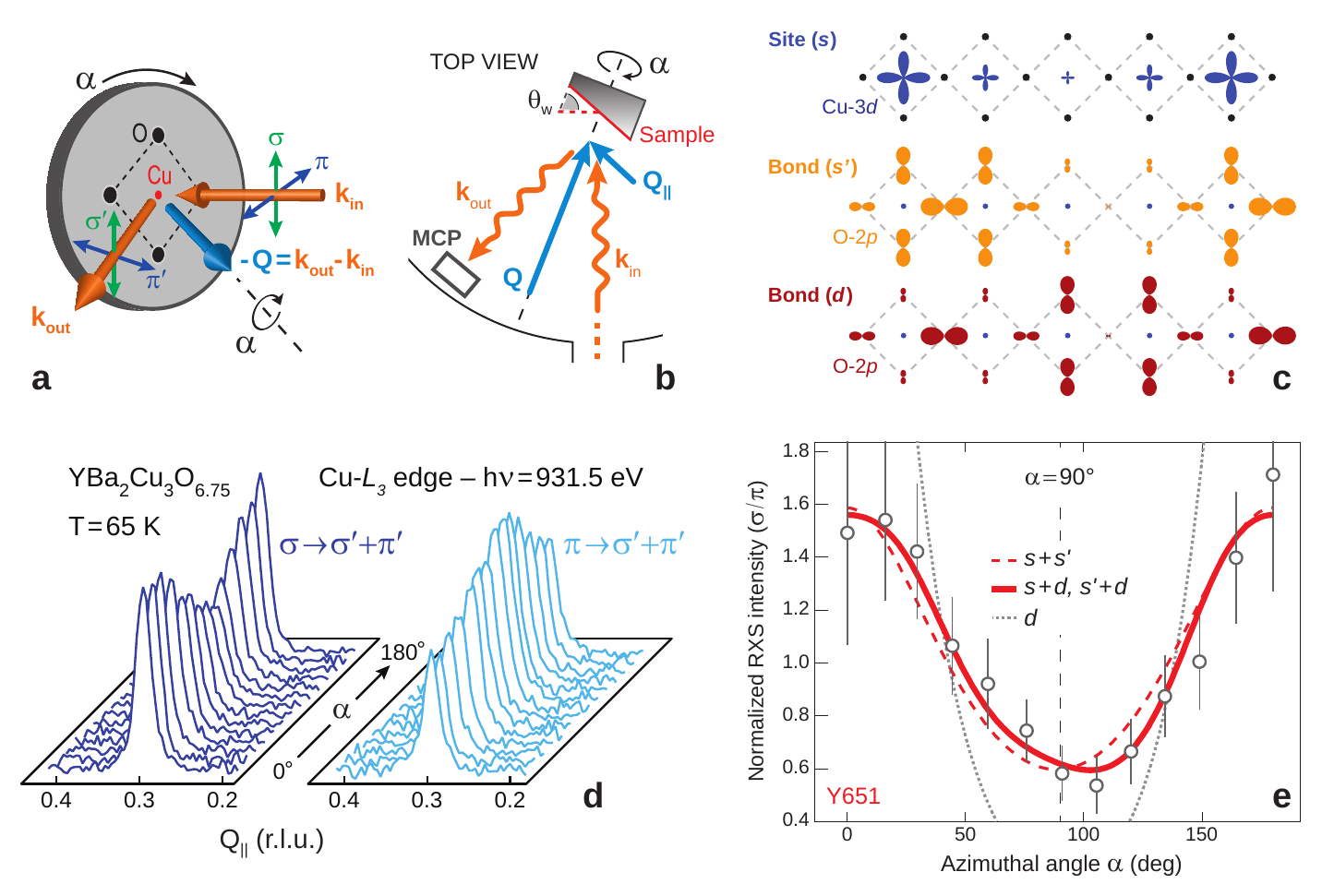}
\caption{\textbf{Azimuthal angle-dependent charge scattering and symmetry of charge order in YBCO.} (\textit{a,b}) Azimuthal geometry for RXS experiments, which allows rotating the crystallographic axes of an angle $\alpha$ around the transferred momentum $\mathbf{Q}$, which is consequently preserved in the frame of reference of the sample. (\textit{c}) Different (but not orthogonal) symmetry representations for the charge ordered state in a 3-orbital system (Cu-$3d$, O-${2p}_{x}$, and O-${2p}_{y}$), such as the cuprates. (\textit{d}) The modulation of the RXS signal vs. azimuthal angle $\alpha$ is evident from the raw RXS scans across the charge order peak in underdoped YBa${}_{2}$Cu${}_{3}$O${}_{6.75}$, for both $\sigma$ and $\pi$ polarizations. (\textit{e}) Azimuthal dependence of the ratio of the $\sigma$ and $\pi$ charge order peak intensities, and comparison to the numerical prediction for different binary combinations of the symmetry terms introduced in (\textit{c}), suggesting a prominent $d$-wave component. Readapted from Ref.\,\citen{Comin_d_wave}.}\label{RSXS_symmetry_YBCO_fig}
\end{figure}
\begin{figure}[t!]
\includegraphics[width=0.9\linewidth]{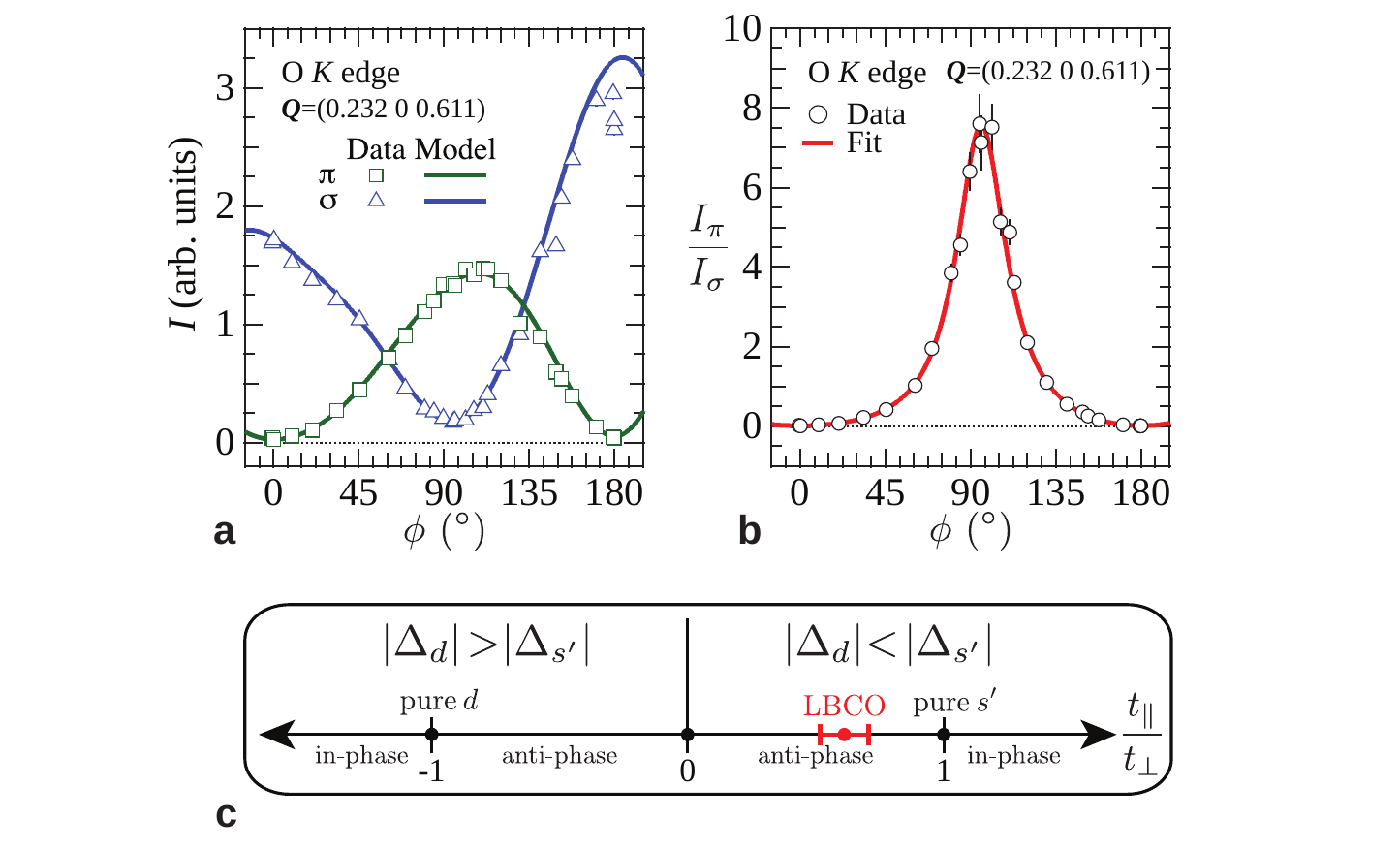}
\caption{\textbf{The symmetry of charge order in LBCO.} (\textit{a}) Azimuthal dependence of the RXS signal from a stripe-ordered LBCO sample, for the two incoming light polarizations $\sigma$ and $\pi$, with theoretical fit profiles achieving best agreement in the case of predominant ${s}^{\prime}$ symmetry. (\textit{b}) Ratio ${I}_{\pi} / {I}_{\sigma}$ of the intensities for the two polarizations, and corresponding best fit profile. (\textit{c}) Diagram of the relative weight of ${s}^{\prime}$- and $d$-wave components as a function of the sign and magnitude of the in-plane (${t}_{\parallel}$) to the out-of-plane (${t}_{\perp}$) x-ray transition amplitudes; the red marker and bar indicates the parameter range yielding the best agreement with the data. Readapted from Ref.\,\citen{Achkar_s_wave}.}\label{RSXS_symmetry_LBCO_fig}
\end{figure}
\begin{equation}
{\Delta}_{\mathrm{CDW}} (\mathbf{k},\mathbf{Q}) \!=\! {\Delta}_{s} + {\Delta}_{{s}^{\prime}} (\cos {k}_{x} \!+\! \cos {k}_{y}) + {\Delta}_{d} (\cos {k}_{x} \!-\! \cos {k}_{y})
\label{CDW_OP_definition}
\end{equation}
\noindent
where ${\Delta}_{s}$, ${\Delta}_{{s}^{\prime}}$, and ${\Delta}_{d}$ represent the magnitude of the \textit{s}-, \textit{s'}-, and \textit{d}-wave terms.

Theoretical predictions in the context of the \textit{t}-\textit{J} model \cite{Vojta_2008,Metlitski2010,Davis_Lee_2013,Sachdev2013} and of the spin-fermion model \cite{Efetov_NatPhys,Chubukov2014} concluded that a \textit{d}-wave pattern of electronic charges is energetically favoured over the other terms. In order to test these theoretical scenarios, an alternative RXS scheme has been recently devised \cite{Comin_d_wave,Achkar_s_wave} and applied to the study of the local symmetry of charge order in Bi2201 and YBCO \cite{Comin_d_wave}, and YBCO and LBCO \cite{Achkar_s_wave}. This approach relies on the definition of a local form factor ${f}_{pq}$ (which is a tensorial quantity, see Eq.\,\ref{form_factor}) capable of discriminating between the different symmetry components of charge order. Once the magnitudes of the different components are built into the local form factor [${f}_{pq} \rightarrow {f}_{pq} \left( {\Delta}_{s}, {\Delta}_{{s}^{\prime}}, {\Delta}_{d} \right) $], the scattering yield can be written as:
\begin{equation}
{I}^{\mathrm{RXS}} \left( \mathbf{Q} \right) \! \propto \! {\left\vert {\sum}_{p q} {\left( {\boldsymbol \varepsilon}_{{\nu}_{\mathrm{in}}} \right)}_{p} \cdot \left[ {\sum}_{n} \: {f}_{pq}^{(n)} \left( {\Delta}_{s}, {\Delta}_{{s}^{\prime}}, {\Delta}_{d} \right) \cdot {e}^{i \mathbf{Q} {\textstyle\cdot} \: {\mathbf{R}}_{n}} \right] \cdot {\left( {\boldsymbol \varepsilon}_{{\nu}_{\mathrm{out}}} \right)}_{q} \right\vert}^{2}
\label{RSXS_azimuthal}
\end{equation}
\noindent
where ${\boldsymbol \varepsilon}_{{\nu}_{\mathrm{in}}}$ and ${\boldsymbol \varepsilon}_{{\nu}_{\mathrm{out}}}$ represent the polarization vectors for incoming and outgoing photons, respectively, while $ \mathbf{{Q}^{*}} $ is the ordering wavevector.
At this point, a single measurement of the RXS intensity will not suffice to resolve the ${\Delta}_{s}, {\Delta}_{{s}^{\prime}}, {\Delta}_{d}$ terms. This issue is overcome by performing the RXS measurement as a function of the sample rotation around the azimuthal axis (see schematic of experimental geometry in Fig.\,\ref{RSXS_symmetry_YBCO_fig}a and b), which is collinear with the ordering wavevector. This procedure allows modulating the projection of the form factor onto the light polarization as a function of the azimuthal angle $\alpha$. Following this approach, the RXS scans can be measured for a range of azimuthal values (see Fig.\,\ref{RSXS_symmetry_YBCO_fig}d for the case of YBCO, from Ref.\,\citen{Comin_d_wave}). The resulting RXS intensities, shown in Fig.\,\ref{RSXS_symmetry_YBCO_fig}e for the case of YBCO at the Cu-${L}_{3}$ edge \cite{Comin_d_wave} and in Fig.\,\ref{RSXS_symmetry_LBCO_fig}a and b for the case of LBCO at the O-$K$ edge \cite{Achkar_s_wave}, can be fitted to the theoretical model of Eq.\ref{RSXS_azimuthal} to evaluate the magnitudes of the symmetry components (which are treated as variational parameters). From these studies, it was concluded that YBCO possesses a prominent \textit{d}-wave symmetry (Fig.\,\ref{RSXS_symmetry_YBCO_fig}e and Ref.\,\citen{Comin_d_wave}), while the stripe order in LBCO is best described by a \textit{s'}-wave pattern (Fig.\,\ref{RSXS_symmetry_LBCO_fig}c and Ref.\,\citen{Achkar_s_wave}). The determination of the charge order symmetry in Bi2201 was instead not conclusive using RXS \cite{Comin_d_wave}. While these early RXS results are poised to stimulate further work to identify and classify these types of symmetries in momentum space, in a recent STM study of Bi2212 and Na-CCOC, Fujita \textit{et al.} \cite{Fujita_2014} have successfully implemented the decomposition of Eq.\,\ref{CDW_OP_definition} by analyzing the reciprocal space representation of the different local symmetries in real space. In particular, these authors first noted how a \textit{d}-wave [\textit{s'}-wave] symmetry suppresses the Fourier amplitudes of the $(\pm Q,0)$ and $(0, \pm Q)$ [$(\pm 1 \pm Q,0)$, $(0,\pm 1 \pm Q)$, $(\pm 1,\pm Q)$, and $(\pm Q,\pm 1)$] peaks, and subsequently were able to resolve the extinction of the inner charge order peaks in reciprocal space starting from the spatially-resolved, partial orbital occupation of the ${O}_{x}$, and ${O}_{y}$ sites. The detailed analysis of the STM conductance maps finally revealed that charge order in Bi2212 and Na-CCOC also possess a predominant $d$-wave form factor.

Another key aspect regarding the microscopic description of charge order, and one that has been debated for long time \cite{kivelson1998,Kivelson_RMP,Norman2004,DelMaestro2006,Robertson2006,Berg2009,Vojta2012,Kivelson_2013}, is whether charge order has a checkerboard (bidirectional) or stripe (unidirectional) character. This problem found an early answer in the La-based cuprates, thanks to the coexistence of spin and charge order with a precise wavevector relation that rules out a checkerboard state. However, magnetic and charge order tend to avoid each other in the phase diagram of all other cuprates, a circumstance which, combined with the typical observation of orthogonal [$(Q,0)$ and $(0,Q)$] charge order reflections, has hindered a conclusive resolution of the checkerboard/stripe dichotomy. Furthermore, evidence in real space using local probes (STM) has been traditionally hampered by the disorder characterizing Bi-based compounds, known to blur the distinction between native uni- and bi-directional ordering instabilities \cite{DelMaestro2006,Robertson2006}; however, recent advancements in the analysis of STM datasets have enabled the visualization of a predominant unidirectional character of electronic modulations in Bi2212 \cite{Hamidian2015}.
\begin{figure}[t!]
\includegraphics[width=0.7\linewidth]{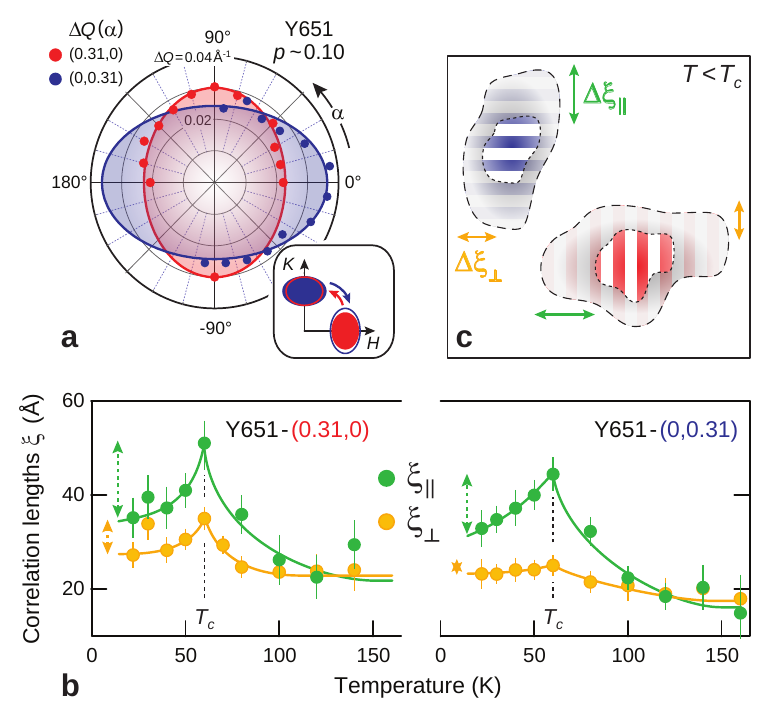}
\caption{\textbf{Stripe vs. checkerboard symmetry of charge modulations in YBCO.} (\textit{a}) Two-dimensional shape of the charge order peaks at $(0.31,0,1.5)$ (red ellipse) and $(0,0.31,1.5)$ (blue ellipse) for a YBa${}_{2}$Cu${}_{3}$O${}_{6.51}$ sample, as determined from fitting the azimuthal-dependent peak widths (red and blue markers). Bottom-right inset: schematic representation of the original peak shapes (full ellipses) and their ${90}^{\circ}$ rotated versions (hollow ellipses). (\textit{b}) Temperature-dependent longitudinal (green) and transverse (yellow) correlation lengths at $(0.31,0,1.5)$ (left) and $(0,0.31,1.5)$ (right), showing a clear anisotropy in the evolution across ${T}_{\mathrm{c}}$. (\textit{c}) Illustration of the anisotropy reported in (\textit{b}), showing how the onset of the superconducting phase reduces the density-density correlations preferentially across the stripes. Readapted from Ref.\,\citen{Comin_stripes}.}\label{RSXS_stripes_YBCO_fig}
\end{figure}

The first indications of unidirectional charge order in underdoped YBCO came from Blackburn \textit{et al.}\,\cite{Blackburn2013} and Blanco-Canosa \textit{et al.}\, \cite{Blanco2013}, showing experimental evidence for very unequal amplitudes between the charge order peak along the \textbf{b} axis (strong) peak and along the \textbf{a} axis (weak)  in YBCO Ortho-II, a direct signature of a tendency to unidirectional behavior in the charge order parameter. 
A recent attempt to assess the character of density modulations was put forth by resolving the full reciprocal-space structure of the charge order peaks in YBCO using RXS \cite{Comin_stripes}. The same geometry was used as outlined in Fig.\,\ref{RSXS_symmetry_YBCO_fig}a and b, which effectively allows to slice through the ordering peak in the $\left( {Q}_{x}, {Q}_{y} \right)$ plane, along different directions in reciprocal space.
In this study, the linewidth from the RXS scans (see again Fig.\,\ref{RSXS_symmetry_YBCO_fig}d) was extracted in order to reconstruct the two-dimensional peak shape as shown in Fig.\,\ref{RSXS_stripes_YBCO_fig}a for a YBa${}_{2}$Cu${}_{3}$O${}_{6.51}$ sample. The elongation of the charge order peaks located along the $(100)$ direction (${\mathbf{Q}}_{a}$) and the $(010)$ direction (${\mathbf{Q}}_{b}$) suggests a locking between the direction of long correlation (narrow peak linewidth) and the wavevector, occurring for the YBa${}_{2}$Cu${}_{3}$O${}_{6.51}$ and YBa${}_{2}$Cu${}_{3}$O${}_{6.67}$ samples -- whereas the charge order peaks in YBa${}_{2}$Cu${}_{3}$O${}_{6.75}$ exhibits the same elongation.
This locking mechanism suggests a breaking of four-fold rotational (${D}_{4h}$) symmetry which is independent for the two ordering components, and is therefore incompatible with a checkerboard state -- the latter being an equal superposition of density modulations along $(100)$ and $(010)$, imposes the same peak structure for the charge order peaks ${\mathbf{Q}}_{a}$ and ${\mathbf{Q}}_{b}$. 

The momentum structure of the charge order peaks already rules out a checkerboard state in favour of a unidirectional (stripy) instability (where stripes can be segregated or overlapping and still generate the same structure in reciprocal space). Further support is provided by the temperature dependence of the longitudinal (${\xi}_{\parallel}$) and transverse (${\xi}_{\perp}$) correlation lengths, which can be derived as the inverse of the peak linewidth along the direction parallel and perpendicular to the wavevector, respectively, and are shown in Fig.\,\ref{RSXS_stripes_YBCO_fig}b. The temperature evolution of the correlation lengths suggests a larger drop, below the superconducting transition temperature ${T}_{\mathrm{c}}$, for the longitudinal (i.e., across the stripes) correlations (see cartoon in Fig.\,\ref{RSXS_stripes_YBCO_fig}c), again leading to a preferential, q-dependent locking of the density-density correlations and to a breaking of fourfold rotational symmetry. 

A similar tendency to a unidirectional character of the charge modulations has also been identified in the different orbital symmetry of the ${\mathbf{Q}}_{a}$ and ${\mathbf{Q}}_{b}$ peaks, measured with RXS \cite{Achkar_s_wave}. This in-plane anisotropy of the short-range charge modulations in YBCO, also detected in NMR measurements \cite{wu2015}, have been argued to cause the large in-plane anisotropy of the Nernst signal seen in YBCO near $p \!=\! 0.12$ \cite{daou2010}, since the Nernst anisotropy grows upon cooling in tandem with the growth in modulation amplitude \cite{CyrChroiniere2015}.

\section{FUTURE PROSPECTS AND NEW CHALLENGES}

\begin{figure}[t!]
\includegraphics[width=0.85\linewidth]{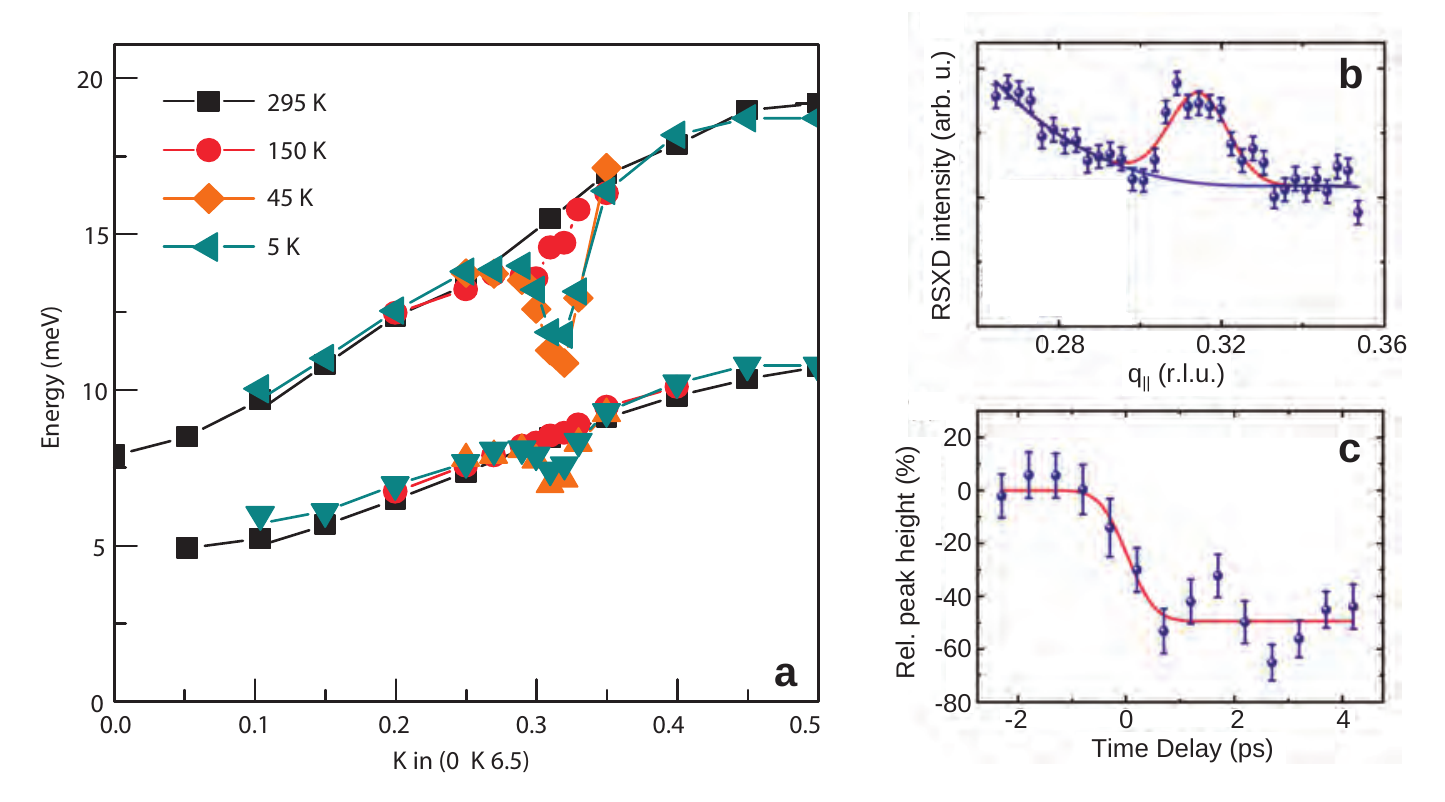}
\vspace*{-4mm}
\caption{\textbf{Giant phonon anomaly and charge order melting from high-resolution frequency- and time-domain x-ray spectroscopy.} (\textit{a}) Momentum and temperature dependence of the frequency of low-energy tranverse acoustic and optical phonons in underdoped YBCO (${T}_{\mathrm{c}} \!=\! 61$\,K), showing a sharp softening occurring at the charge order wavevector ${\mathbf{Q}}_{\mathrm{CO}} \!=\! (0, 0.31, 6.5)$ in the superconducting phase ($T < {T}_{\mathrm{c}}$). Readapted from Ref.\,\citen{LeTacon2013}. (\textit{b}) Charge order peak in underdoped YBCO (${T}_{\mathrm{c}} \!=\! 61$\,K), acquired using ultrafast soft x-ray pulses from the Linear Coherent Light Source. (\textit{c}) Corresponding time-resolved dynamics of the charge order peak amplitude following impulsive photoexcitation ($t \!=\! 0$) of the apical oxygen mode ($\lambda \!\sim\! 15 \mu$m) using 400\,fs pulses. Readapted from Ref.\,\citen{Forst_2014}.}\label{RSXS_new_developments}
\end{figure}

Despite the recent flurry of experimental findings and theoretical insights on charge order in the cuprates, there is still a lot to learn and to understand. First and foremost, the mechanism driving the doped holes into breaking translational symmetry has not been conclusively pinned down. In particular, it is unclear whether the cuprates exhibit Peierls physics similar to other low-dimensional compounds \cite{Gruner_1994} or whether a new and unconventional mechanism is at play. Several studies have been performed since the discovery of stripe order, aimed at elucidating the electron-lattice interplay and its relevance for charge order instabilities \cite{Reznik_2006,Reznik_2007,Reznik_2008,Reznik_review_2010,Raichle_2011,Reznik_2012}. Recently, the work by Le Tacon \textit{et al.} \cite{LeTacon2013} revealed a strong and sharp (in momentum) softening of the low-energy acoustic and optical phonons at the charge order wavevector in underdoped YBCO [see Fig.\,\ref{RSXS_new_developments}a -- similar observations were also made for NCCO \cite{Dastuto_2002} and Bi2201 \cite{Bonnoit2012}]. This result \textit{per se} already provides direct evidence of a pronounced electron-lattice coupling mechanism, while the partial yet incomplete softening (with the frequency remaining nonzero) further clarifies that the charge order in cuprates is not due to a phonon mode freezing into a static lattice distortion, consistent with the short-ranged nature of charge correlations at zero magnetic field. The charge order peak remains confined to the quasi-elastic line, which bears the typical signatures of a `central peak', which is characteristic of the presence of ordered nanodomains which gradually coalesce into a state with longer correlations. However, the most puzzling finding emerges from the temperature dependence of the softening effect, which is found to be enhanced in the superconducting phase, where charge order is paradoxically weakened. This behavior reflects a strong superconductivity-induced phonon anomaly, occurring at the wavevector of the charge instability, thus exposing a very complex intertwining between these competing phases and might deserve further explorations for a recent theoretical discussion of this phenomenon see Ref.\,\citen{Liu2015}).

Time-resolved experimental methods, which probe the recovery dynamics of the system following an ultrafast perturbation, have rapidly emerged as an alternative tool to gain further insights on correlated materials. Depending on the nature of the excitation, pump-probe techniques can probe both the microscopic coupling between different degrees of freedom near equilibrium (i.e., in the linear response regime attained at low excitation intensities) \cite{Orenstein_2012,DalConte_2012,Torchinksy2013,Hinton2013} or explore completely new physics in the strongly perturbative regime (at high excitation intensities) where photoinduced phases can be accessed that are often inaccessible at equilibrium \cite{Fausti_2011}. The field of time-resolved scattering and diffraction has seen rapid advancements since its inception at the end of the 1990's \cite{Rose_Petruck_1999}. New avenues in the x-ray study of ordering phenomena out-of-equilibrium have been enabled by the recent development of free-electron-laser (FEL) sources, which generate bright ultrashort light pulses over a broad range of energies from the extreme ultraviolet to the hard x-rays, thereby creating new opportunities for the investigation of structural and electron dynamics with high time resolution. The spectral control of the photoexcitation process is an additional experimental parameter, enabling the selective perturbation of specific degrees of freedom via the nonlinear coupling between light and lattice \cite{Forst_2011,Mankowsky_2014}. When this framework is combined with photon energy tunability, resonant scattering can be used to probe the recovery of electronic orders that are brought out-of-equilibrium by an ultrafast pump \cite{Forst_2011_LSMO,Caviglia_2013}. In the context of charge order in the cuprates, these measurements were performed for the first time by F\"{o}rst \textit{et al.} \cite{Forst_2014}, using ultrafast THz light pulses to photoexcite a particular lattice vibration (the apical oxygen mode) in underdoped YBCO and soft x-ray FEL pulses (at the Cu-${L}_{3}$) to track the evolution of the charge order peak (Fig.\,\ref{RSXS_new_developments}b) as a function of time. The photoexcitation process strengthens the superconducting state by promoting coherent interlayer transport and the charge order amplitude is correspondingly weakened by a factor two (see Fig.\,\ref{RSXS_new_developments}c), thus providing new insights on the competition between charge order and superconductivity in a regime where phase coherence is artificially enhanced with light \cite{Kaiser_2014,Hu_2014} [similar results were also reported in LBCO \cite{Forst_2014_LBCO}]. More in general, the new FEL capabilities are poised to considerably extend the domain of application of resonant x-ray methods toward the study of nanoscale ordering phenomena at unprecedented length- and time-scales.

Another new frontier for the use of high-brightness x-ray sources is in the context of high-magnetic field studies. The use of high fields within different experimental techniques, such as quantum oscillations or NMR, has been transformative for our understanding of high-temperature superconductors, enabling unprecedented insights into the nature and interplay of competing orders in these complex materials. It was then natural to envisage the extension of high-field capabilities to scattering and diffraction experiments and, notwithstanding the complications caused by the requirement for optical access for incoming and outgoing photon beams, several efforts have been successfully carried forward in this direction in recent years. Chang \textit{et al.} \cite{chang2012} used a superconducting cryomagnet specifically designed for low-angle forward-scattering measurements in transmission geometry and using very penetrating high energy photons ($100$\,keV), which enabled the first observation of magnetic field-induced enhancement of the charge order signal. 
%
%

Very recently, an innovative experimental scheme was introduced by Gerber \textit{et al.} \cite{Gerber2015} that is based on the use of a high-field (28\,T) split-pair pulsed magnet synchronized with the ultra-bright and ultrafast x-ray pulses (with photon energy of 8.8\,keV) of the Linear Coherent Light Source. This approach, leveraging the high single-pulse photon flux of the FEL beam, enabled the acquisition of two-dimensional single-shot diffraction patterns with sufficient photon counts in spite of the very low sampling frequency (with one spectrum every 2 to 25 minutes, due to the recovery time of the pulsed magnetic field apparatus). The momentum-space maps of charge order in YBCO, from 0 to 25\,T, reveal an ostensible evolution in the momentum structure of the charge order peak as a function of both the in-plane ($K$) and out-of-plane ($L$) wavevectors. In particular, the out-of-plane character of the charge order peak changes from a very broad, elongated structure at zero field to a well-defined peak (with $\sim\! 170$\,\AA\, correlation length) located around $L \!=\! 1$. This is accompanied by a sharpening of the in-plane peak linewidth, with a corresponding increase of charge correlations from $\sim\! 60$ to $\sim\! 180$\,\AA, and a concomitant enhancement of the diffracted intensity. These results disclose the complexity of charge order and its rich and unconventional phenomenology as a function of doping as well as magnetic field. In particular, this study not only uncovers the momentum-resolved crossover between the low- and high-field regimes, previously investigated using NMR \cite{wu2015}, but also clarifies that a full reconstruction of the Fermi surface as seen by quantum oscillations only happens in high field when long range order sets in, thereby elucidating the absence of folding and small pockets in the ARPES measurements (which can only be performed in zero field) on YBCO \cite{Hossain,FournierNP}.

To conclude, we would like to emphasize that the further development of these novel approaches and methodologies will enable the exploration of completely new dimensions. This will bring us many more surprises and much deeper insights in the study of symmetry breaking instabilities in cuprates, and will lead to a fuller understanding of these phenomena and of their intimate interplay with high-temperature superconductivity.

\begin{summary}[SUMMARY POINTS]
\begin{enumerate}
\item \textbf{Technical advancements}: The challenge posed by the detection and characterization of charge-density waves in the context of the rich phenomenology of the doped CuO${}_{2}$ planes has propelled tremendous advancements in the field of soft x-ray scattering methods. Presently, RXS beamlines are operational or under construction at several facilities worldwide: ALS, APS, BESSY, CLS, DESY, Diamond, ESRF, NSLS-II, NSRRC, SLS, SOLEIL, Spring-8, SSRL.
\item \textbf{Universality}: To-date, evidence of charge order has been found for all hole-doped cuprate families, as well as in Nd-based electron-doped compounds, using a variety of complementary experimental probes.
\item \textbf{Resolution}: Resonant x-ray scattering has been successfully used to detect charge density modulations with a spatial coherence as short as 15-20\,\AA, thus extending to momentum space capabilities that were previously accessible only to spatially-resolved techniques (STM).
\item \textbf{Symmetry}: The richness of the RXS information, and the multiple control parameters available -- polarization, photon energy, sample orientation -- have opened up the possibility to explore new microscopic aspects of the ordered state -- such as its dimensionality (1D vs. 2D) and its local symmetry (\textit{s}-, \textit{s'}-, and \textit{d}-wave).
\end{enumerate}
\end{summary}

\begin{issues}[FUTURE DIRECTIONS]
\begin{enumerate}
\item Unveiling the driving force behind charge order in cuprates.
\item Find new approaches to modulate and, in general, control charge order.
\item Elucidate the relative role of Mott physics and the proximity to quantum-critical behavior for charge order, as well as superconductivity, in cuprates.
\item Probe the evolution of charge order and its interplay with superconductivity under extreme conditions -- high magnetic fields, high pressures, ultrafast optical pumping.
\end{enumerate}
\end{issues}

\section*{DISCLOSURE STATEMENT}
The authors are not aware of any affiliations, memberships, funding, or financial holdings that might be perceived as affecting the objectivity of this review.

\section*{ACKNOWLEDGMENTS}
The authors are indebted to P. Abbamonte, L. Braicovich, A. Cavalleri, E. H. da Silva Neto, J. C. Davis, J. Fink, J. Geck, G. Ghiringhelli, C. Giannetti, R. L. Greene, M. Greven, D. G. Hawthorn, F. He, J. E. Hoffman, M.-H. Julien, B. Keimer, M. Le Tacon, P. A. Lee, W. S. Lee, C. Mazzoli, M. R. Norman, S. Sachdev, G. A. Sawatzky, R. Sutarto, L. Taillefer, and J. M. Tranquada for valuable discussions and for the critical reading of this review.

This work was supported by the Max Planck-UBC Centre for Quantum Materials, the Killam, A. P. Sloan, A. von Humboldt, and NSERC's Steacie Memorial Fellowships (A.D.), the Canada Research Chairs Program (A.D.), NSERC, CFI and CIFAR Quantum Materials.


\providecommand{\noopsort}[1]{}\providecommand{\singleletter}[1]{#1}%

\end{document}